\documentclass{PoS}

\title{Experimental Evidence of Black Holes}

\ShortTitle{Experimental Evidence of Black Holes}

\author{\speaker{Andreas M\"uller}\\ 
        Max--Planck--Institut f\"ur extraterrestrische Physik,  p.o. box 1312, D--85741 Garching, Germany\\
        E-mail: \email{amueller@mpe.mpg.de}}

\abstract{Classical black holes are solutions of the field equations of General Relativity. Many astronomical 
observations suggest that black holes really exist in nature. However, an unambiguous proof for their 
existence is still lacking. Neither event horizon nor intrinsic curvature singularity have been observed by 
means of astronomical techniques. \\
This paper introduces to particular features of black holes. Then, we give a synopsis on current 
astronomical techniques to detect black holes. Further methods are outlined that will become important in the 
near future. For the first time, the zoo of black hole detection techniques is completely presented and 
classified into kinematical, spectro--relativistic, accretive, eruptive, obscurative, aberrative, temporal, 
and gravitational--wave induced verification methods. 
Principal and technical obstacles avoid undoubtfully proving black hole existence. We critically discuss 
alternatives to the black hole. However, classical rotating Kerr black holes are still the best theoretical 
model to explain astronomical observations.}

\FullConference{School on Particle Physics, Gravity and Cosmology\\
		21 August - 2 September 2006 \\
		Dubrovnik, Croatia}

\begin{document}

\section{Introduction} 
Black holes (BHs) are the most compact objects known in the Universe. They are the most efficient gravitational lens, 
a lens that captures even light. Albert Einstein's General Relativity (GR) is a powerful theory to describe BHs mathematically.
It is a theory that describes Gravity as curved space--time. In a sense, Gravity is not a force but a geometric feature
of a deformable 4D continuum. According to this picture, BHs are an extreme form of curved space--time containing a curvature
singularity that swallows matter and light.

Astonishingly, at the advent of GR in 1916 
the first solution of Einstein's complicated field equation was found in the same year. This (external) Schwarzschild solution 
\cite{Schwarzschild1916} describes the simplest BH type mathematically. However, the notion \textit{black hole} was coined 
significantly later, in 1967 by the relativist John Archibald Wheeler. 

It soon became clear that nature may allow for the existence of these mysterious objects. In the classical GR collapse of 
massive stars nothing can stop gravitational pressure so that a BH forms. This picture was essentially pushed forward by the 
work of Julius Robert Oppenheimer and Hartland Snyder in 1939 \cite{Oppenheimer1939}. The result of these developments is 
the first BH mass family: \textbf{stellar--mass BHs}. 

In the Sixties the understanding of active galactic nuclei (AGN) like quasars and radio galaxies delivered another piece to 
the BH puzzle \cite{Zeldovich1964, Salpeter1964, LyndenBell1969, LyndenBell1971}. The extreme quasar luminosities of 
$10^{47}$ erg/s $\sim 10^{14}\,{\rm L}_\odot$\footnote{${\rm L}_\odot\simeq 3.85\times 10^{33}$ erg/s} could be explained by 
BHs that swallow matter. Accretion turned out to be the most efficient process to produce radiation in the whole range of 
electromagnetic emission. However, estimates with the Eddington relation \cite{Eddington1924} pointed to BHs in the centers 
of galaxies that are significantly more massive than the first family. This second mass family is called \textbf{supermassive 
black holes} (SMBHs).
Paradoxically, the darkest object in the Universe produces the most powerful source of luminosity. Of course,
this mechanism works outside the event horizon that marks the ultimate blackness. Nowadays, astrophysicists are convinced
that stellar--mass BHs play the key role in black hole X--ray binaries (BHXBs) such as microquasars, in gamma--ray bursts 
(GRBs), and maybe also in ultra--luminous X--ray sources (ULXs) \cite{Fabbiano1989, Colbert1999}. Recently, 
it has been shown that the timescale of X--ray variations from stellar and supermassive BHs are in fact physically linked 
if one corrects for the accretion rate \cite{McHardy2006}. The same physics is at work.

The giant SMBHs are supposed to inhabit almost any center of galaxies causing an active quasars--like phase in an earlier 
state of the galaxy evolution. Local galaxies contain non--active or dormant SMBHs. 
 
Currently, there is an active debate on the cosmological role of BHs. In hierarchical galaxy formation scenarios is it possible 
to explain luminous quasars at cosmological redshift $z\sim 6$ \cite{Volonteri2003}. SMBHs are formed on cosmological timescales
from seed BHs with several hundred solar masses. Probably, these seeds originated from the first population of very massive 
stars. The initial BHs grew by merging and accretion to form the nearby SMBHs. In the same scenarios it is possible to explain
the spin--up of BHs in this context of BH growth. The spin distribution is dominated by rapidly spinning BHs -- essentially by 
gas accretion. Maximally spinning BHs are the result of thin disk accretion but similarly for thick disk accretion about 4/5 
have very high spins, $a\sim 0.8 \ M$ \cite{Volonteri2005}. 

Another interesting finding is the so--called luminosity---dependent density evolution. Data from ROSAT, XMM--Newton, and 
Chandra surveys have been collected together to provide huge samples with $\sim 1000$ AGN \cite{Hasinger2005}. Based on these
data it could be shown that the number density of high--luminosity AGN ('quasars') peaks at significantly lower cosmological 
redshift than for low--luminosity AGN ('Seyfert galaxies'). This anti--hierarchical growth can be explained theoretically
\cite{Merloni2004}.

Currently, astronomers use multi--wavelength surveys to collect a huge amount of observational data. The analysis 
reveals interesting connections between galaxy and central SMBH. There seems to be a cosmological link between the two that 
points towards co--evolution scenarios. 

A thorough introduction to BH physics and BH phenomenology has been given by Jorge Zanelli and Neven Bili\' c at this school, 
see \cite{Bilic2006b} and references therein. Another review can be found in \cite{Mueller2004b}\footnote{Download at
\scriptsize{\texttt{http://www.mpe.mpg.de/\~\,amueller/downloads/PhD/PhD\_AMueller.pdf}}}

This contribution is devoted to experimental techniques in astronomy that are used to detect cosmic BHs. After an introduction 
into BH features a complete presentation and classification of detection methods is given. We have a closer look at BH spin 
measurements. Finally, we critically discuss the question what we observe: What are the astronomical facts? What are possible 
alternatives to the classical BH solutions? 

\section{Black hole features} \label{sec:bhfeat}
In the present paper, the treatment is restricted to classical BHs i.e.\ the 
(external) Schwarzschild solution \cite{Schwarzschild1916} and the Kerr solution \cite{Kerr1963}.
Essentially, this is justified by two reasons: first, space--time rotation is a crucial ingredient 
for BH astrophysics e.g.\ in the view of paradigms for relativistic jets launching in AGN \cite{Blandford1977} 
and GRBs \cite{Meszaros1997, Meszaros2002}. 
There is still no rotating BH alternative available to date that could replace a
classical Kerr BH. Second, the regime of classical GR is supposed to be a good choice in 
some distance to a BH candidate. Deviations from GR solutions due to quantum gravity effects are expected to 
emerge at length scales that are comparable to the Schwarzschild radius or even smaller. This is currently 
'out of astronomical range'. The regime of strong gravity has already clear signatures starting at 
some distance to the horizon and getting stronger as approaching the hole. As suggested by relativistic 
emission line diagnostics this critical distance amounts to several tens gravitational radii 
\cite{Mueller2004a}. Hence, the restriction to classical GR BH solutions is sufficient. 

The natural length scale of GR is the gravitational radius defined by $r_{\rm g}={\rm G}M/{\rm c}^2$ with 
Newton's constant ${\rm G}$ and vacuum speed of light ${\rm c}$. For one solar mass, 
$M={\rm M_\odot}=1.989\times 10^{33}\,{\rm g}$, the Schwarzschild radius amounts to $R_{\rm S}\simeq 3\,{\rm km}$. 
In theory, it is convenient to use geometrized units ${\rm G}={\rm c}=1$ so that length is measured in units of 
mass $M$ \cite{Misner1973}. We measure BH rotation in terms of the specific angular momentum (\textit{Kerr parameter}) 
that holds $a=J/M=\mathrm{G}M/\mathrm{c}$ \cite{Cunningham1973}. At first glance, the Kerr parameter can take any value 
between $-M$ and $+M$ but the cases $a=\pm M$ develop a naked singularity that is forbidden \cite{Penrose1969}. Accretion 
theory favours a limit of $|a|_{\rm max}=0.998\,M$ \cite{Thorne1974}.

To prepare to detection techniques for BHs, we first arrange the properties of 
these objects in the next two subsections.  
%

\subsection{Blackness} \label{sec:black}
The blackness of BHs is a result of strongly curved space--time forcing light rays (null geodesics) 
to bend back to the gravitational source. This is a consequence of their compactness. A suitable 
parameter to measure compactness of a stellar body with mass $M_\ast$ and radius $r_\ast$ is the 
dimensionless quantity \cite{Mueller2004b}
\begin{equation} \label{eq:comp}
C\equiv GM_\ast/(c^2r_\ast), 
\end{equation}
which is denoted \textit{compactness parameter} hereafter. 

$C$ varies between zero and unity where a value close to unity signals an extremely compact object.
Let us look at some examples: White dwarfs at the Chandrasekhar limit may satisfy $C_\mathrm{WD}\simeq 0.0004$, 
neutron stars hold $C_\mathrm{NS}\simeq 0.16$, quark stars may have $C_\mathrm{QS}\simeq 0.37$, 
Schwarzschild BHs\footnote{Here, the event horizon radius is assumed to be the 'surface' of a BH.} hold 
$C_\mathrm{SBH}=1/2$ and extreme Kerr BHs satisfy $C_\mathrm{KBH}=1$.
Therefore, BHs rotating at their limit\footnote{$a=M$ is assumed here for simplicity but the exact value 
of the maximum angular momentum of BHs is still under debate. Usually, one assumes Thorne's limit 
$a\simeq 0.998\,M$ \cite{Thorne1974}, see however \cite{Aschenbach2004b, Shapiro2005} for alternatives.} 
represent the most compact objects in the Universe. 

In the framework of classical GR a point mass is described by the (external) Schwarzschild solution. 
Schwarzschild BHs represent a one--parameter family in GR. The static Schwarzschild space--time is
fully described by the mass parameter $M$. 

GR tells us that mass causes gravitational redshift effects i.e.\ the strong pull of gravity shifts radiation 
emitted near masses towards the red spectral range and the intensity is reduced, too. This is mathematically described 
by a general relativistic Doppler factor or $g$--factor for short hereafter \cite{Mueller2004a}. The $g$--factor is a 
measure for both, energy shift of emission and suppression or enhancement of intensity as compared to the emitter's 
rest frame. $g$ is generally a function of the velocity field of the emitter (as measured in a suitable observer's frame 
e.g.\ the zero angular momentum observer, ZAMO), the curved space--time (represented by metric coefficients), and 
constants of motion of the photon as discovered in \cite{Carter1968}. One important metric coefficient is the 
redshift or lapse function $\alpha$ that measures the amount of gravitational redshift of photons and time 
dilation. It holds for static BHs 
\begin{equation} \label{eq:lapse_S}
\alpha_\mathrm{S} = \sqrt{1-\frac{2M}{r}},
\end{equation}
and is known as \textit{Schwarzschild factor}. In a good approximation, any slowly rotating mass may be 
described by this factor. 

In contrast, Kerr BHs belong to a two--parameter family described by two physical quantities, 
mass $M$, and specific angular momentum $a$ \cite{Bardeen1974}. 
%
%
%
%
%
%
%
In Boyer--Lindquist form \cite{Boyer1967}, a Kerr black hole is given by the line element\footnote{The signature 
of the metric is $(-+++)$ throughout the paper.}
\begin{equation} \label{eq:BLKerr}
ds^{2}=-\alpha^{2}dt^{2}+\tilde\omega^{2}(d\phi-\omega dt)^{2}+\rho^{2}/\Delta \ dr^{2}+\rho^{2} d\theta^{2},
\end{equation}
where expression (\ref{eq:lapse_S}) can be generalized to be the lapse for rotating BHs
\begin{equation} \label{eq:lapse_K}
\alpha = \frac{\rho\sqrt{\Delta}}{\Sigma}.
\end{equation}
Other metric functions satisfy 
\begin{eqnarray}
\Delta & = & r^{2}-2Mr+a^{2}, \\
\rho^{2} & = & r^{2}+a^{2}\cos^{2}\theta, \\
\Sigma^{2} & = & (r^{2}+a^{2})^{2}-a^{2}\Delta\sin^{2}\theta, \\
\omega & = & 2aMr/\Sigma^2, \label{eq:omeg} \\
\tilde{\omega} & = & \Sigma\sin\theta/\rho. \label{eq:omegtild}
\end{eqnarray}

Per definition, any classical BH horizon satifies $\Delta=0$ and consequently 
$\alpha_\mathrm{S}=\alpha=0$. Since $g\propto\alpha$, the redshift influences 
\textit{any emission} of electromagnetic radiation in the vicinity of a BH. Observed 
intensity of radiation satisfies $F^\mathrm{obs}_\nu \propto g^3$ so that spectral flux is strongly 
reduced as the emitter approaches to the event horizon. The consequence is the main BH feature: 
\textit{darkness} and even \textit{blackness} at the event horizon. For static Schwarzschild BHs, 
$a=0$, there is only one horizon that is located at the Schwarzschild radius, $\mathrm{R_S}=2M$. 
For rotating Kerr BHs there are in fact two horizons that hold 
\begin{equation} \label{eq:horiz}
r^\pm_\mathrm{H}=M\pm\sqrt{M^2-a^2},
\end{equation}
where $r^+_\mathrm{H}$ denotes the outer horizon and $r^-_\mathrm{H}$ the inner horizon. The inner 
or Cauchy horizon can only be intersected once i.e.\ it is a one--way ticket to the curvature 
singularity. Following the cosmic censorship conjecture by Roger Penrose the intrinsic 'true' 
singularity is hidden by these horizons \cite{Penrose1969}.

We conclude that the essential BH indicator is darkness as a result from gravitational redshift. 
Hence, astronomers seek for 'black spots' (BS) in the sky especially at positions where BH candidates are supposed 
to be e.g.\ in X--ray binaries (XRBs), gamma--ray burst remnants (GRBRs), possibly in centres of globular clusters 
and confidently in centres of galaxies. The technical challenge is to
resolve these tiny spots. The apparent size $\theta_\mathrm{BH}$ of the next SMBH in the centre of 
the Milky Way can be found on the micro--arcsecond ($\mu$as) scale! We provide a useful equation to compute 
this immediately from BH mass $M$ and BH distance $d$: 
\begin{eqnarray}\label{eq:app_bh_size}
\theta_\mathrm{BH} & = & 2\,\arctan\left(R_\mathrm{S}/d\right) \nonumber\\
& \simeq & 2\,R_\mathrm{S}/d \nonumber\\
& \simeq & 39.4\times\left(\frac{M}{10^6\,{\rm M}_\odot}\right)\times\left(\frac{1\,\mathrm{kpc}}{d}\right)\,\mu\mathrm{as}.
\end{eqnarray}
The values are adapted to SMBHs that have typical masses that range from million to billion solar masses and 
typical distances from several kpc to Mpc\footnote{$1 \ {\rm pc}=3.26 \ {\rm lightyears}\simeq 3.1\times 10^{18} \ {\rm cm}$}.  
It is sufficient to consider only the first order term of the $\arctan$ expansion since $R_\mathrm{S}/d\ll 1$ 
for cosmic BH candidates. \\
\subsection{Rotating space--time}
Kerr BHs reveal a speciality that lacks in case of Schwarzschild BHs: rotation of space--time. The spin of Kerr 
BHs forces anything to co--rotate: matter, light, close observers, because space itself rotates. The 
forcing property is called \textit{frame--dragging}. The frame--dragging frequency already introduced as 
$\omega$ in Eq. (\ref{eq:omeg}), parameterizes the rotation of space as viewed from infinity. Approaching the 
Kerr BH, the 'spin of space' steeply increases, $\omega\propto\,r^{-3}$. It is important to note that 
Schwarzschild BHs force anything \textit{not} to rotate, $\omega=0$, a feature that may be entitled as 
\textit{anti--frame--dragging}. 
  
Kerr BHs are endowed with a zone where the rotation of space--time becomes extraordinarily strong. It called the 
ergoregion that is oblate. The outer edge of the ergoregion, the ergosphere, can be computed according to $g_\mathrm{tt}=0$ 
and we find
\begin{equation}
r_\mathrm{erg}(\vartheta)=M+\sqrt{M^2-a^2\cos^2\vartheta}.
\end{equation}
The metric coefficient flips sign at this radius: it is negative for $r>r_\mathrm{erg}$ and positive for
$r<r_\mathrm{erg}$. In the equatorial plane, $\vartheta=\pi/2$, the ergosphere lies at the Schwarzschild radius -- independent 
of BH spin. However, for lower poloidal angles the ergosphere approaches the BH horizon and coincides with 
it at the poles, $\vartheta=0$.

Frame--dragging can be nicely interpreted by an analogue to electromagnetism, called gravitomagnetism \cite{Thorne1986}.
From this view, a Kerr BH interacts by gravitoelectric and gravitomagnetic forces with its surroundings.
Gravitoelectric forces are just ordinary gravitational forces attracting test bodies. But the interpretation
of gravitomagnetic forces is more complicated: One relevant phenomenon associated with gravitomagnetic forces 
is the Lense--Thirring effect. 
It simply states that two gyroscopes interact by exchanging gravitomagnetic forces. One realisation of two gyroscopes 
is a rotating BH surrounded by a rotating accretion disk. In this case the Kerr BH pushes the rotating accretion disk 
into the equatorial plane by means of gravitomagnetic forces \cite{Bardeen1975}. The transition region can be found at 
$10^2$ to $10^4$ gravitational radii.
The accretion disk whobbles around the BH. Hence, hot disk radiation that is detected in the X--ray range might 
be variable and show quasi--periodic features in the power density spectra that are dominated by the Lense--Thirring 
frequency \cite{vanderKlis2000}. 

BH rotation is important: Rapidly rotating space--time plus magnetic fields are the main ingredients to drive 
relativistic jets magnetohydrodynamically. Recently, numerical non--radiative general relativistic magnetohydrodynamics 
(GRMHD) simulations have shown that Kerr BHs rotating near their limit are efficient rotators to spin--up magnetic 
field lines and drive Poynting fluxes \cite{Koide2002, Krolik2005, McKinney2006}. 
Schwarzschild BHs are not sufficient to explain the observed relativistic jets of AGN and GRBs. 
%

\begin{figure}[!t]
	\begin{center}
	\rotatebox{0}{\includegraphics[width=.6\textwidth]{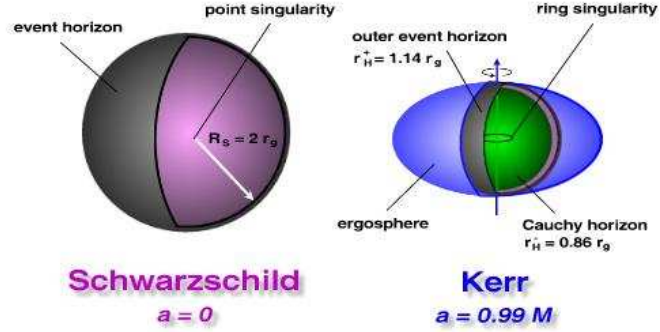}}
	\caption{Architecture of static Schwarzschild ({\em left}) and rotating Kerr BHs ({\em right}). We assume 
	the same mass. One immediately notices that Kerr BHs are more compact than Schwarzschild BHs.} \label{fig:BHs}
	\end{center}
\end{figure}

Finally, Fig.~\ref{fig:BHs} gives a comparison on BHs and compares static ({\em left}) with rotating BHs 
({\em right}). We stress that this is just a 'structural view' showing the main BH features. But it is 
\textit{not} an invariant view because it is based on the radii of a specific coordinate system. A better 
and invariant view onto BHs is possible by using curvature invariants such as the Kretschmann scalar 
\cite{Henry2000} -- however, this is not that intuitive.

We summarize this section that blackness and rotating space--time are the two observable properties
of classical black holes. 

\section{Black hole detection methods} \label{sec:detec}
How can one detect a cosmic BH? In principle, one can distinguish direct and indirect methods. Direct 
evidences for cosmic BHs means that an astronomer has to prove the key features of a classical black hole by
observations: the event horizon and the curvature singularity. Due to the cosmic censorship conjecture by
Roger Penrose \cite{Penrose1969} BH intrinsic singularities are hidden by an event horizon. Therefore, a proof 
of the curvature singularity seems to be impossible. The proof of an event horizon means to show strong evidence 
for a zero--emission region because at event horizons the general relativistic Doppler factor vanishes exactly, 
$g=0$, due to zero redshift (or lapse) $\alpha=0$. On one hand, this is certainly difficult because any astronomical 
brightness measurements involve error bars so that only a $g\approx 0$--statement could be made at best. On the 
other hand, Hawking has demonstrated that in a semi--classical quantum gravity description (quantized fields on a 
GR background metric) black hole event horizons radiate \cite{Hawking1975}. Based on these arguments, it seems 
\textit{principally} impossible to prove direct evidence for cosmic BHs by means of electromagnetic radiation. 
We feel that the only possibility for direct methods consists in clear signatures of \textit{gravitational waves} 
as pointed out recently \cite{Abramowicz2002, Berti2006}. 

Hence, the diversity of current methods could be summarized as indirect methods. Astronomers measure the amount of 
mass and the associated volume. From density arguments one may arrive at the plausible conclusion that nothing else 
can fit but a BH. Of course, astronomers have to test thouroughly if alternative scenarios may fit such as compact 
stellar clusters or other types of compact objects, e.g.\ neutron stars, boson stars, or fermion balls etc. 
If not, a good BH candidate is detected.

In the following, we present different methods that are in use by astronomers. But we also go beyond practice and 
show further methods that might be applied in the near future. All these methods are classified by a suitable label.

\subsection{Kinematical methods}
%
Kinematical methods to detect black holes are widely--used and very successful. The simple idea is that moving 
objects that feel the deep gravitational potential of the black hole serve as tracers to derive properties of 
the black hole i.e.\ mass and spin. Hence, it is an \textit{indirect} method because the observer
does not care about the black hole itself, but about the tracer. In practise suitable tracers are luminous 
stars, gas or flaring objects. Any test body that is luminous enough to be detected on Earth can in principle
serve as dynamical tracer. Typically the tracer surrounds the BH on Keplerian orbits. In that case, the tracking
time scale is determined by the Keplerian time scale at given radius.

\paragraph{Pioneering work} The first BH was detected by the Canadian astronomer Tom Bolton in 1971 \cite{Bolton1972}. 
He studied the X--ray binary system Cygnus X--1. Bolton was able to measure the radial velocities of the giant 
O--star HDE 226868. Therefore, he determined the mass function by a kinematical method. The compact 
companion was favoured to be a stellar BH because alternatives such as a white dwarf or a neutron star 
were ruled out. Today, the BH in Cygnus X--1 amounts to $\sim 10\,{\rm M}_\odot$ and HDE 226868 has 
$\sim 18\,{\rm M}_\odot$. The orbital period amounts to 5.6 days. This is the first historical BH verification.

\begin{figure}[!t]\begin{center}
	\rotatebox{0}{\includegraphics[width=8.0cm]{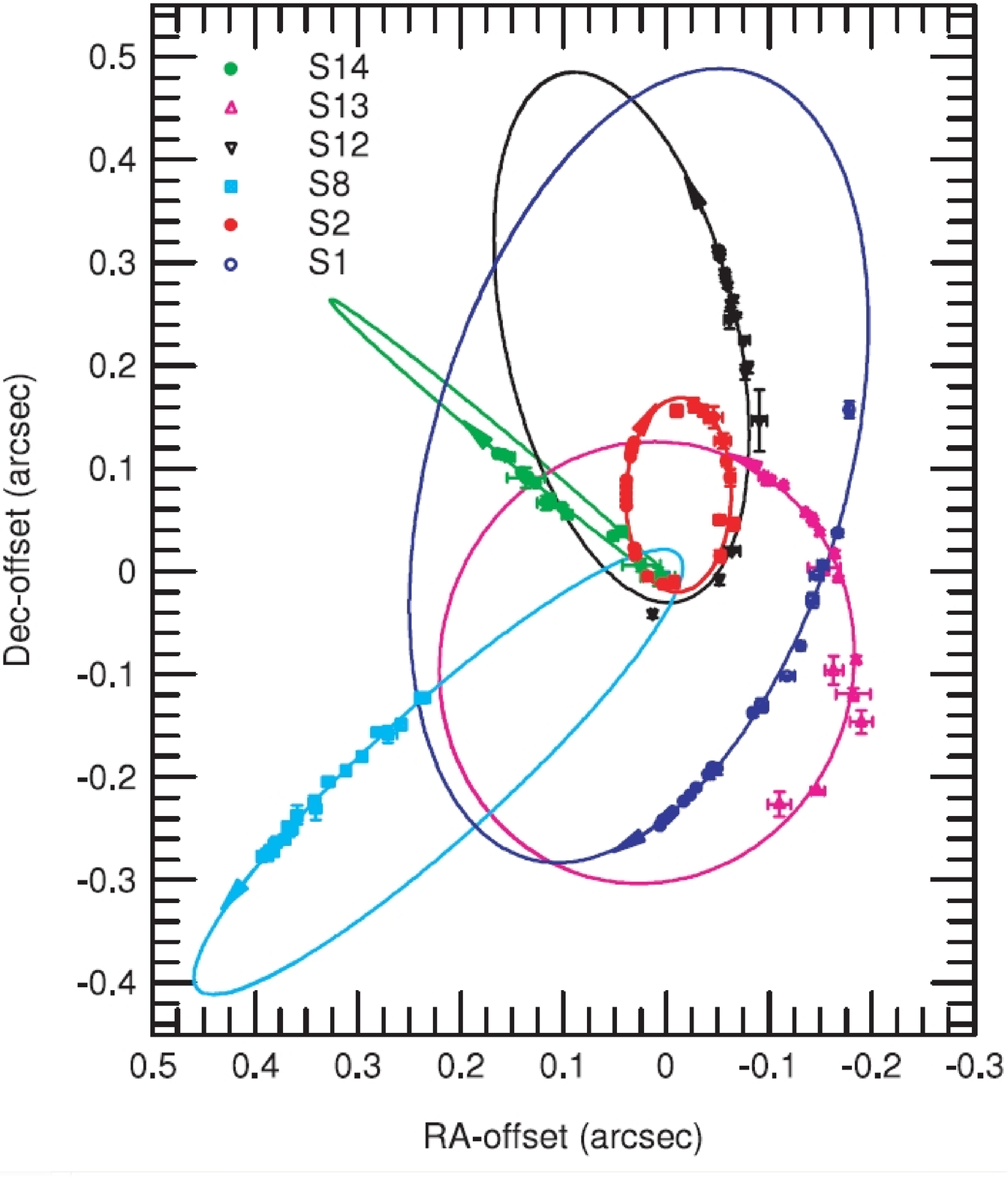}}
 	\caption{Observed orbital shapes of six S--stars orbiting the Galactic 
	Centre SMBH as projected to the sky. The image is based on NIR observations 
	taken from \cite{Eisenhauer2005}.} \label{fig:GCorb}
\end{center}\end{figure}

\paragraph{Single stellar orbits} Another example is the compact radio source Sgr A* that is associated with the SMBH 
in the Galactic Centre (GC). Unfortunately, the view is obscured in the optical waveband so that astronomers have to use 
radio waves, near--infrared (NIR) radiation or X--rays to look into the centre of the Milky Way.
The first infrared group started in 1992 to  monitor gas and motions of stars around the GC BH 
\cite{Eckart1992, Eckart1996}. Essentially, these results are confirmed by another group \cite{Ghez1998, Ghez2003}.
%
%
%
%
%
%
Proper motion and radial velocity measurements reveal Keplerian orbits of the S--stars allowing to 
determine a periastron distance of 16 light hours for the star \textit{S2}. Fig.~\ref{fig:GCorb} shows
nicely the Kepler ellipses of this innermost star and other S--stars. From this the distance of the 
Galactic Centre is updated to be $R_0=7.62\pm 0.32$ kpc. From Kepler's third law a central mass of 
$(3.61\pm 0.32)\times 10^6 \ {\rm M}_\odot$ is deduced \cite{Eisenhauer2005, Paumard2005}.
%
%
%
%
%

This compact and dark mass is very likely a SMBH because any other alternatives such 
as compact star clusters, boson stars, or fermion balls are ruled out \cite{Genzel1997, Schoedel2002}. 
Stars orbiting SMBHs are thereby suitable tracers to determine BH features. Astronomers plan follow--up 
projects to approach the BH in our Galaxy and to probe GR effects (see Sec.~\ref{sec:aberr})
%
%
%
%
%
%
However, in theoretical units the star \textit{S2} is still far away from the BH because the 
periastron distance amounts $\approx 1600 \ \mathrm{R_S}$. This is rather the asymptotical flat 
region of space--time, $\alpha_\mathrm{S}\,(r=1600 \ \mathrm{R_S})\simeq 0.9997$. Strong gravity 
effects become important at a few tens Schwarzschild radii as proposed by relativistic emission 
line diagnostics \cite{Mueller2004a}. 
%
%

\paragraph{$M$--$\sigma$ relation} But not only single stars are suited. Ensembles of stars that follow 
their orbits in groups can also be used e.g.\ if the resolution is not sufficient enough to detect 
single stars. This situation is typical for distant galaxies. Slit spectroscopy of galactic nuclei is a 
valuable tool to analyse stellar velocity dispersion profiles. The study of galaxy samples shows evidence 
for a strong correlation between mass of the central BH, $M$, and the stellar velocity dispersion, 
$\sigma$, the prominent M--$\sigma$ relation \cite{Gebhardt2000}. Plotted as log--linear relation 
\begin{equation}
\log(M/M_\odot)=\alpha+\beta \, \log(\sigma/\sigma_0),
\end{equation}
with a suited reference value chosen to $\sigma_0=200$ km/s this correlation can be fitted to 
galaxy samples. 
Recent results fit the slope to $\beta=4.0\pm 0.3$ \cite{Tremaine2002}. This value undershoots 
significantly first fits of $\beta=5.27\pm 0.4$ \cite{Ferrarese2000} and overshoots another one 
$\beta=3.75\pm 0.3$ \cite{Gebhardt2000}. Previously, another relation was found that links BH 
mass and bulge mass \cite{Magorrian1998}. 
%
%
Both correlations hint for a physical connection between stars from the galactic bulge and the 
central BH. There is a controversial debate if Seyferts and quasars follow the same 
$M$--$M_\mathrm{bulge}$ relation \cite{McLure2001} or not \cite{Wandel1999}.

The M--$\sigma$ relation is also theoretically understood \cite{Haehnelt2000, Adams2001}. 
%
%
However, very recent results from M31 (including Milky Way and M32) state that 
the $M$--$\sigma$ relation has significant intrinsic scatter -- at least at low BH 
masses \cite{Bender2005}.
The redshift evolution of the $M$--$\sigma$ relation was probed by using quasars 
\cite{Shields2003}. Whereas faint normal galaxies drop out of observability, these luminous AGN 
can still be detected at high cosmological redshifts. BH masses are measured from
continuum luminosities and width of H$\beta$ lines. Narrow 
$[\mathrm{OIII}]$ line widths serve as tracers to determine the velocity dispersions. Out to
$z\simeq 3$ the $M$--$\sigma$ relation is still complied. However, at very high redshifts, 
there are observational hints that the strong correlation, $M\propto\sigma^4$, breaks down. 
Recently, it was suggested to use a curved M--$\sigma$ relation \cite{Wyithe2006}.

\paragraph{Reverberation mapping} Another kinematical verification method of black holes involves the 
emission of gas. It is called reverberation mapping technique. The broad line regions (BLRs) are supposed 
to be luminous matter clouds almost consisting of hydrogen that surrounds a galactic core. Due to their 
fast motion emission lines are significantly broadened by the Doppler effect. Therefore, the line width 
is a measure for the velocity of the BLRs, $\sigma$. Another essential parameter involved in reverberation 
mapping is the distance of the BLRs to the centre of that galaxy, $r$. This is determined by comparing 
primary emission from the galactic core (continuum radiation) with secondary emission from the BLRs 
(emission lines) which is nothing else than the response of the core radiation. From the Virial theorem 
the central mass is deduced to be
\begin{equation}
M\approx r\sigma^2/\mathrm{G}.
\end{equation}
Today, astronomers are convinced that in nearly any center of a galaxy there is a SMBH. An active i.e.\ 
accreting SMBH thereby causes the enormous AGN luminosity. In the local Universe the SMBHs in galaxies 
are dormant and inactive e.g.\ Sgr A* or M31*. But if there are luminous orbiting clouds available such as 
BLRs (especially for AGN type--1 i.e.\ AGN that allow observers to look into the core), then there is good 
chance to measure the BH mass. 

\paragraph{Maser emission} The next kinematical method that is based on gas motion involves maser line 
emission. Water masers can be found in the molecular torus located at the pc scale of a galaxy. This 
coherent microwave emission from Keplerian rotating gas orbiting the SMBH can be used to measure the 
central mass. In case of NGC~4258, the BH mass could be determined to $3.6\times10^7\,{\rm M}_\odot$ within 
0.13 pc \cite{Pietsch2002} and in case NGC~3079 astronomers found $2\times 10^6\,{\rm M}_\odot$ enclosed within 
0.4 pc \cite{Kondratko2004}.

\paragraph{Kinematical BH spin measurements} It is challenging to determine whether or not a BH rotates 
by using kinematical techniques. In most cases indicators are to far away from the BH. Due to 
$\omega\propto r^{-3}$ behaviour, rotation of space--time is extraordinarily strong only at a few gravitational 
radii distance to the BH. 

\paragraph{Quasi--periodic flare emission} close to the GC 
BH has been observed in NIR and in X--rays: The NIR flare exhibits a quasi--periodicity of $\sim$17 min 
that is interpreted as relativistically modulated emission of circulating gas. Assuming that the flare 
emitter moves at last stable Keplerian circular orbit (= innermost stable circular orbit, ISCO) the observed 
period matches to a Kerr parameter for the GC BH to be $a\simeq 0.52\pm 0.26$ \cite{Genzel2003}. 
The power density spectra of the X--ray flares observed at the GC BH revealed 
distinct peaks yielding periods of $\sim$100 s, 219 s, 700 s, 1150 s, and 2250 s. A comparison
with characteristic frequencies associated with accretion disks i.e.\ Lense--Thirring precession, 
Keplerian orbital, vertical, and radial epicyclic frequency, match to a BH satisfying 
$M=2.72\times 10^6\,{\rm M}_\odot$ and $a=0.9939\,M$  \cite{Aschenbach2004a}.
The nature of the flare emitters -- possibly linked to accretion flow or orbiting stars -- is still 
unknown.

\subsection{Spectro--relativistic methods} \label{sec:spec}
In general, the notion spectro--relativistic refers to relativistic effects in observed spectra that can be used
to determine BH properties. The condition for this idea to work is of course that the emitter has to be sufficiently 
close to the BH. 

\begin{figure}[!t]
	\begin{center}
	\rotatebox{0}{\includegraphics[width=.6\textwidth]{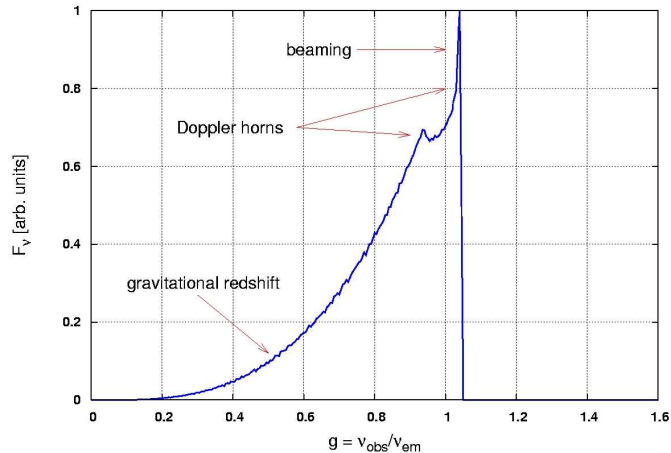}}
	\caption{Prototype of an observed relativistic emission line profile that is emitted from a thin accretion 
	disk around a Kerr BH with $a/M = 0.998$. The disk extends from 1.24 (ISCO) to 100 gravitational radii and 
	is inclined to $30^\circ$ (typical value for AGN type--1). The radial disk emissivity profile obeys 
	$\epsilon(r)\propto r^{-3}$. The line profile is subject of three effects: The Doppler effect due to rotation 
	of the disk causes two horns; special relativistic beaming intensifies the blue line wing, and gravitational 
	redshift smears out the whole line profile and produces an extended red line wing.} \label{fig:line}
	\end{center}
\end{figure}

\paragraph{Fe K lines} Let us assume a very sharp line profile in the rest frame of the emitter e.g.\ an accretion disk.
Further, this disk rotates around a BH and extends down to the ISCO i.e.\ only a few gravitational radii away from
the event horizon. An astronomical observer in a distant laboratory frame would observe a line profile that is very 
different from that in the rest frame because the line photons are subject to a number of physical effects that happen
on their way to the observer. First, the Doppler effect produces a Doppler redshift at the part that is receding and
a Doppler blueshift at the part of the disk that is approaching to the observer. The result is a broadened line maybe
with two Doppler peaks if the disk is sufficiently inclined towards the observer (longitudinal Doppler effect). This 
already happens in Newtonian physics. Second, for intermediate to large inclinations there is a special relativistic 
beaming effect at work. The Keplerian velocities are indeed comparable to the speed of light so that radiation is bent
towards the observer. Therefore, the observed radiation intensity is amplified as compared to the rest frame. Third, the
BH captures line photons by strong space--time curvature. This gravitational redshift effect is two--fold as described 
in Sec.~\ref{sec:black}: the observed photon energy is redshifted relative to the rest frame energy and the spectral 
line flux is lower in the observer's frame. All these three effects produce an asymmetric and skewed relativistic emission
line profile that is shown in Fig.~\ref{fig:line}.
X--ray astronomers often observe such kind of profiles as iron K fluorescent lines \cite{Fabian2000}. They are part 
of the so--called 'reflection bump' in X--ray spectra that occurs around several tens of keV. This bump is the response 
of the ionized accretion disk when hit by hard X--ray radiation e.g.\ from the primary Compton continuum. The most prominent 
iron fluorescent line is the neutral Fe K$\alpha$ line that has a rest frame line energy of 6.4 keV. In 1995, the first 
relativistically broadened iron K profile has been observed in the Seyfert--1 galaxy MCG--6--30--15 with the Japanese 
X--ray observatory ASCA \cite{Tanaka1995}. It was found that even a Kerr BH would be consistent with these observations.

Meanwhile, many examples are observed that exhibit this kind of relativistic iron lines. This includes quasars 
\cite{Mineo2000, Porquet2004} and stellar--mass BHs (see \cite{Miller2006b} for a recent review). 

\begin{figure}[!t]
	\begin{center}
	\rotatebox{0}{\includegraphics[width=.6\textwidth]{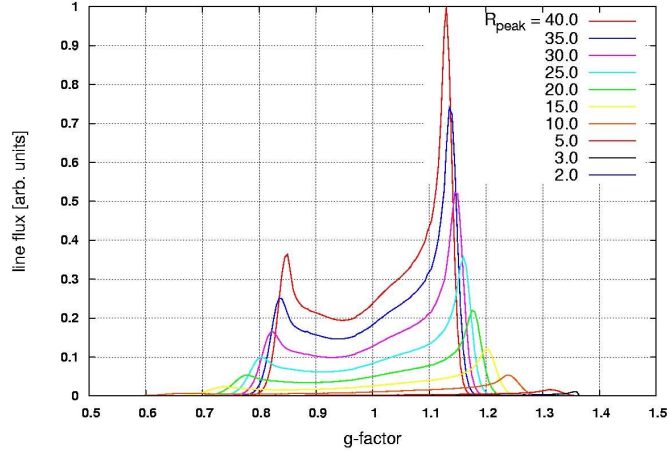}}
	\caption{Relativistic emission line profiles from rings in Keplerian rotation around a Kerr BH obeying 
	$a/M = 0.998$. The rings are approximately one gravitational radius narrow and inclined to $75^\circ$. 
	A Gaussian emissivity profile was chosen to mimic luminous rings (see \cite{Mueller2004a} for details). 
	The $g$--factor is just the full GR Doppler factor that is defined by $g=\nu_{\rm obs}/\nu_{\rm em}$ 
	with radiation frequency in the observer's frame $\nu_{\rm obs}$ and emitter's frame $\nu_{\rm em}$. 
	$R_{\rm peak}$ is the ring radius with maxium emission and the only variable quantity in this study.
	The emission line profiles become broader and fainter as the rings move towards the BH.} \label{fig:emlines}
	\end{center}
\end{figure}

In principle, gravitational redshift is a long--ranging effect that extends to infinity. However, the effect
decreases steeply with distance and is hard to detect in some distance to the mass or BH. In a recent theoretical 
study it was possible to show that gravitational redshift can be probed out to several 10000 gravitational radii 
distance to the BH in AGN if sufficient spectral resolution with modern telescopes is imposed \cite{Mueller2006}. 
Indeed, this is feasible today and was demonstrated for the Narrow--Line Seyfert--1 galaxy Mrk~110 in the optical 
\cite{Kollatschny2003} and in X--rays \cite{Boller2006}. This offers a new detection method for BHs to derive masses
with multi--wavelength data. All gravitationally redshifted features should point towards the same mass. 
Fig.~\ref{fig:emlines} is based on ray tracing simulation in the Kerr geometry and shows how the line profile 
develops when approaching the BH \cite{Mueller2006}. Essentially, the BH space--time induces a relativistic line 
broadening and a line suppression in the vicinity of the BH. 

\begin{figure}[!t]
	\begin{center}
	\rotatebox{0}{\includegraphics[width=.6\textwidth]{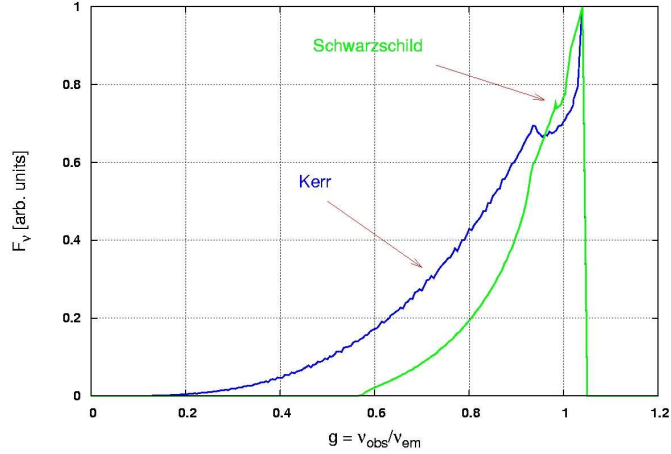}}
	\caption{Direct comparison of observed relativistic emission line profiles around a Kerr ($a/M = 0.998$, 
	blue) and a Schwarzschild BH ($a/M = 0$, green). The inner disk edge is identical to the ISCO in each case i.e.\
	$r_{\rm in}=1.24\,r_{\rm g}$ (Kerr) and $r_{\rm in}=6.0\,r_{\rm g}$ (Schwarzschild). All other parameters are the 
	same in both cases such as disk inclination angle $i=30^\circ$, disk outer edge $r_{\rm out}=100.0\,r_{\rm g}$ and 
	single power law emissivity $\epsilon(r)\propto r^{-3}$. The velocity field of the emitting disk is Keplerian.} \label{fig:KSlines}
	\end{center}
\end{figure}

Relativistic emission line profiles also indicate BH spin. The basic idea is the following: It is known from 
accretion theory that standard disks extend down to the ISCO \cite{Shakura1973}. The ISCO depends on BH spin and 
is closer to the event horizon for rapidly spinning Kerr BHs, $r_{\rm ISCO}(a=M)=1\,r_{\rm g}$, as compared to 
Schwarzschild BHs, $r_{\rm ISCO}(a=0)=6\,r_{\rm g}$. As a consequence the line emission comes from regions that
are deeper in the gravitational potential if the BH rotates. Therefore, the line profiles from the vicinity of 
Kerr BHs are more influenced by gravitational redshift. This is documented in Fig.~\ref{fig:KSlines} where a profile
from a Kerr BH is compared to that of a Schwarzschild BH and the inner disk edge couples to the ISCO in each case 
(all other line parameters are the same). As expected, the line shape is broader and exhibits an extended broad 
red wing for the Kerr BH whereas the line shape around the Schwarzschild BH is narrow. The red wing serves as a 
tracer for the inner disk edge and (due to the imposed coupling) for BH spin.

However, this idea is plagued by the fact that real accretion disks may \textit{not} extend down to the ISCO. 
Radiative cooling might cause disk truncation and produce a narrow profile even though the BH rotates 
\cite{Mueller2004a}. In any case, an observed broad profile strongly favours a Kerr BH. But the often imposed 
coupling of ISCO and Kerr parameter should be handled with care.

\paragraph{Multitemperature blackbody spectrum} The rest frame spectrum that is emitted near the BH could be of other 
kind, too. Thin accretion disks are known to have a temperature profile and a temperature maximum close to the
inner disk edge \cite{Shakura1973}. Let us divide the disk into rings each with a specific temperature. Then, the 
resulting disk blackbody spectrum is built--up by the blackbody of each ring. This spectrum is called multitemperature 
blackbody spectrum. If the disk is close enough to the BH GR effects are expected to influence the blackbody radiation.
In fact, this is the case and a ray tracing model has been proposed \cite{Li2005}. These authors also compared this
model to observations of BHXBs and were able to measure BH spins. They found rather high Kerr parameters, $a/M\gtrsim 0.7$.

\subsection{Accretive methods}
%
%
%
Accretive methods involve active BHs i.e.\ BHs that accrete matter
thereby causing luminous effects. As mentioned in the introduction there is in principle
the same physics involved in both cases, stellar and supermassive BHs \cite{McHardy2006}.

\begin{figure}[!t]
	\begin{center}
	\rotatebox{0}{\includegraphics[width=.6\textwidth]{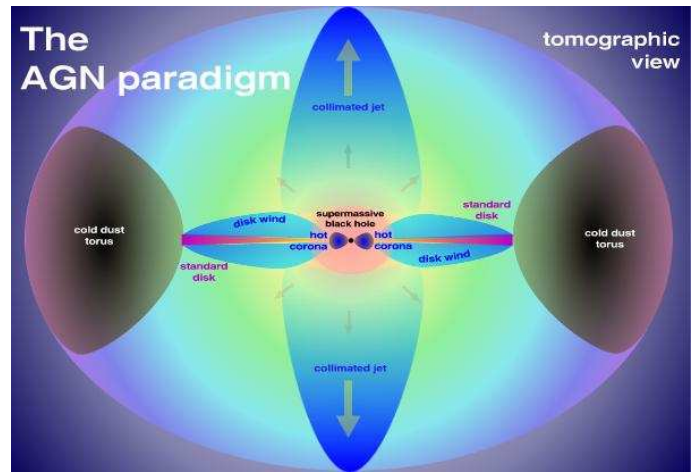}}
	\caption{Sketch of the inner parts of an active galactic nucleus (AGN). Consider that this is a 
	sectional view with logarithmic scale along the axes. The main structures in an AGN are dusty torus,
	standard accretion disk, hot advection--dominated inner accretion flow, central SMBH, and Jets.} \label{fig:AGN}
	\end{center}
\end{figure}

\paragraph{AGN paradigm} First, we turn to the AGN paradigm. Fig.~\ref{fig:AGN} gives a sketch overview:
At the sub--parsec scale there is a huge
rotating donut--like configuration, the cold dusty torus. This matter flows due to instabilities 
into the center of the AGN. Thereby, a geometrically flat and optically thick accretion flow develops:
the standard accretion disk. Depending mainly on accretion rate a geometrically thick and optically 
thin advection--dominated accretion flow (ADAF) \cite{Narayan1994} may form only several tens of 
gravitational radii away from the central SMBH. In the vicinity of the SMBH magnetic effects are 
favoured to drive strong outflows, the Jets, that are observed in radio galaxies and (radio--loud) 
quasars. Therefore, any AGN discovery proves an active SMBH.

\paragraph{Jet base investigations} The physics in the direct SMBH vicinity is currently explored in theory.
The theoretical branch is called general relativistic magnetohydrodynamics (GRMHD). There are a number
of groups world--wide that have GRMHD codes to simulate the interaction of accretion flow, BH and Jet 
\cite{Koide1999, DeVillier2002, Gammie2003, Duez2005, Anninos2005, Anton2006}\footnote{Here, the first 
papers of each group are given; there are many follow--up papers available.}. The time evolution of 
different initial configurations is studied in these supercomputer simulations e.g.\ plasma tori or 
accretion disks or just an electromagnetic field. In 1991 it turned out that weak magnetic fields in a 
differentially rotating configuration are the essential input for an operating MHD instability, the
magnetorotational instability (MRI) \cite{Balbus1991}. The MRI drives strong magnetic turbulence thereby 
producing efficient angular momentum transport. This is the precondition to start accretion. MHD turbulence
pushes an inflow into the BH. Then, two mechanisms may produce magnetically the jets. The first mechanism 
involves magnetic flux tubes that pierce the accretion disk. Analoguously to the solar surface the flux 
tubes extract matter from the disk causing a magnetically driven disk wind that may form the jets after 
collimation. This extraction mechanism not involving a BH is known as Blandford--Payne scenario \cite{Blandford1982}. 
Here, the energy source is the magnetized accretion flow. The second mechanism needs a \textit{rotating BH}. 
The accretion flow transports the field lines into the BH's ergosphere. Frame--dragging spins up the magnetic 
field lines and produces a strong magnetic field. At some point the field 
is strong enough to produce leptonic pairs. An outward directed electromagnetic energy flux (Poynting flux)
accelerates the plasma. The associated MHD waves drive the plasma magnetically. This is called the 
Blandford--Znajek process \cite{Blandford1977}. The energy source is the rotational energy of the Kerr BH. 
Currently, the two mechanisms are under debate. 

Radio astronomers hope two clarify this point because very long baseline interferometry 
with millimetre radio waves (mm--VLBI) already has spatial resolutions that amount to only 15--20 Schwarzschild 
radii for near AGN such as M87 in Virgo \cite{Krichbaum2006a}. mm--VLBI offers the best resolution available to 
date. The structures at the jet base will give rise for the jet launching mechanism at work. If the jet base
is very small in size e.g.\ only a few Schwarzschild radii this would favour the Blandford--Znajek process and 
disfavour the Blandford--Payne scenario because the latter involves an extended disk with at least several tens
of Schwarzschild radii in size \cite{Krichbaum2005}. Further, if astronomers compare the observed structures 
with GRMHD simulations this might constrain BH mass and spin. Mass is restricted by the apparent size of the BH 
at the expected position and also by the apparent spatial distance of the jet bases if a two--sided jet is visible 
e.g.\ in case of Cyg A \cite{Krichbaum2006b}.

\paragraph{Eddington's relation} Using Eddington's relation \cite{Eddington1924} that 
links luminosity to accretion rate and BH mass, accretive methods are of particular interest
for high accretion rates close to the Eddington rate. 
Accretion onto a SMBH is a widely accepted paradigm that drives AGN luminosity. But also 
X--ray binaries that contain stellar black holes are observed where black hole accretion 
drives a strong X--ray source. Prominent examples are Cyg X--1 and Cyg X--3. Recently, for the
BHXB GRO J1655-40 it could be shown observationally that magnetic accretion operates as expected from
the Blandford--Payne scenario \cite{Miller2006a}. Thereby, the different accretion and spectral states 
of BHXBs can be described in a unified scheme \cite{Esin1997}. From the observed luminosity it is possible 
to estimate the BH mass. A higher value of the radiative efficiency $\epsilon\simeq 0.1...0.4$ even indicates 
BH spin. 

\paragraph{Deep field observations} The final method that could be termed as accretive verification method
involves deep field observations i.e.\ astronomical observations with very long exposure times. Then, very 
faint and distant objects can be studied. X--ray deep field observations of the Lockman hole \cite{Hasinger2004} 
suggest that $\sim 100$ AGN in that field can be combined together to form a summed feature at $\sim 6.4$
keV rest frame energy. This is interpreted as a stacked Fe K$\alpha$ feature \cite{Streblyanska2005}. 
The best fitting line profile suggests rotating BHs. This finding is consistent with another analysis 
\cite{Brusa2005}.

\subsection{Eruptive methods} \label{sec:eru}
Eruptive verification methods are always associated with burst--like phenomena. Such events are quite common in astronomy 
and the challenge is to select those that are truly associated with a BH.

\paragraph{Gamma--ray bursts} The best examples are GRBs. Due to current paradigms for short--term \cite{Narayan1992, Ruffert1995} 
and for long--term GRBs \cite{MacFadyen1999} astrophysicists are convinced that in \textit{each} GRB there forms a stellar--mass BH. 
Short--term GRBs are merging binaries consisting of two neutron stars or a neutron star and a BH. Long--term GRBs are also called 
hypernovae because they exceed the total energy output of supernovae by a factor of $\sim 100$. A gravitational collapse of a massive
star e.g.\ a Wolf--Rayet star leads to a hypernova. In any case a stellar--mass BH forms and drives ultrarelativistic jets. They
produce a prompt $\gamma$--ray emission that is followed by X--ray, optical, and radio afterglows, see \cite{vanParadijs2000} for a 
review.

\begin{figure}[!t]
	\begin{center}
	\rotatebox{0}{\includegraphics[width=.6\textwidth]{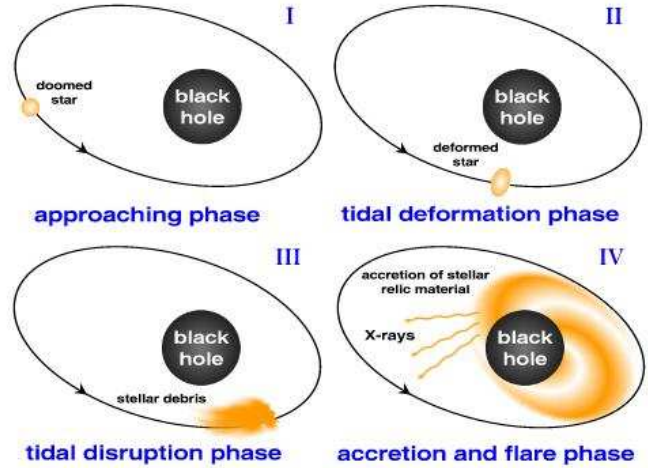}}
	\caption{Sketch of the tidal disruption scenario, see text for details.} \label{fig:disrup}
	\end{center}
\end{figure}

\paragraph{Tidal disruption events} Another eruptive detection method for BHs involves stellar tidal disruptions. Fig.~\ref{fig:disrup} 
illustrates this scenario: In the approaching phase (I) the star with mass $m_\ast$ and radius $R_\ast$ approaches a BH of mass $M$.
In the tidal deformation phase (II) the star is strongly deformed by tidal forces of the BH. At the tidal radius that satisfies    
\begin{equation}
R_{\rm T}\approx R_\ast\left(M/m_\ast\right)^{1/3}
\end{equation}
the BH's tidal forces dominate selfgravity of the star. As a consequence the star is disrupted in the tidal disruption phase (III).
The stellar debris spread over a region close to the BH and might be accreted partly. If so, a characteristic X--ray flare
develops in the accretion and flare phase (IV). 

This X--ray flare has typical signatures that hint for a tidal disruption e.g.\ it is thermal radiation, the X--ray luminosity decays
with time according to a power law, $L_{\rm X}\propto t^{-5/3}$ \cite{Evans1989} with typical timescales of months and the integrated 
luminosity yields an energy output of $\simeq 10^{51}$ erg that is comparable to supernovae. Stellar tidal disruptions are rare events 
that occur once in $10^4$ years per galaxy \cite{Magorrian1999}. To exclude flares from AGN activity the X--ray observations have to be 
accompanied by simultaneous optical observations.

The observation of a stellar tidal disruption of a solar--type star constrains immediately the BH mass to be smaller than $\approx 10^8\,M_\odot$
because for larger BHs the tidal radius is smaller that the Schwarzschild radius \cite{Hills1975}.
 
In 2004, it has been reported on a stellar tidal disruption that occured in the non--active galaxy RX J1242-1119 \cite{Komossa2004, Halpern2004}. 
A huge drop of the X--ray luminosity in the post--flare spectrum by a factor of $\simeq 200$ and other characteristics as mentioned before 
favour that indeed a stellar disruption has been detected. The brightness in the blue band as observed with the Hubble space telescope gives
an estimate for the SMBH mass to $\simeq 2\times 10^8\,{\rm M_\odot}$.

Recently, another stellar tidal disruption event has been reported also in the UV \cite{Gezari2006}. The BH mass can be derived from blackbody 
spectral fits. Due to the fact that the radius of emission should be greater than ISCO these fits also provide an estimate of BH spin.

\paragraph{Hawking evaporation} 
In the early Seventies, it was shown that even a semi--classical approach with quantized scalar fields on the 
background of non--quantized 4D BH space--times results in the possibility that BHs \textit{are not} totally black 
\cite{Hawking1975}. Practically, this Hawking emission is to dim for astrophysical BH candidates and estimates 
show that the horizon illuminated by Hawking radiation deviates hardly from the classical zero--luminosity horizon. 
On one hand, the signal--to--noise ratio does not allow to detect dim Hawking emission currently; 
on the other hand, BH surroundings are blurred with other kinds of radiation that are even more
intense.

For typical cosmic BHs the decay due to energy loss by Hawking radiation can be neglected because the decay time
scales with $\tau\propto M^{-3}$. A solar--mass BH decays after $10^{64}$ years exceeding the Hubble time by 
far. However, Hawking evaporation might be interesting for primordial BHs that have canonical masses around 
$10^{15}$ g \cite{Carr2003} -- but still there are no observational clues that hint for these small BHs in the Early
Universe.

We make an aside and concentrate onto a new branch of BH physics that will be of great importance in the near future 
that is to say BHs in particle accelerators. Possibly, high--energy physics will deliver fruitful new insights into
BH research: On the condition that spatial extra dimensions (more than three) exist in nature and the classical 
Planck scale is reduced to significantly lower energy scales, it may be possible to produce mini BHs in high--energy 
particle collisions \cite{Dimopoulos2001, Giddings2002, Cavaglia2003}. 

The general formula to compute the reduced Planck scale, $M_\mathrm{Pl,D}$, from the number
of spatial extra dimensions, $n$, and the compactification Radius, $R$, is in the ADD scenario 
\cite{ArkaniHamed1998}
\begin{equation}
M_\mathrm{Pl,D}=M_\mathrm{Pl}^{2/(2+n)}\left(\frac{\hbar}{Rc}\right)^{n/(2+n)},
\end{equation}
where $M_\mathrm{Pl}$ denotes the classical Planck scale of $M_\mathrm{Pl}=\sqrt{\hbar c/G}=1.221\times 10^{19}$ GeV,
$\hbar$ Planck's constant, $c$ vacuum speed of light and $G$ Newton's constant. For all cases $n\gtrsim 2$ (as suggested 
by M--theory, string theories and supergravity) a modified gravity occurs at the sub--mm scale. This theory is called TeV 
quantum gravity because the Planck energy is reduced and comparable to the electroweak scale, $m_{\rm EW}\sim 1$ TeV.  
However, the mm to $\mu$m regime has been tested but spatial extra dimensions have not been found so far. TeV quantum gravity 
scenarios can be tested at the Large Hadron Collider (LHC) at CERN or at International Linear Collider (ILC). 
Maybe, it will be possible to verify Hawking emission in terrestrial experiments because evaporating mini BHs are produced
in in TeV quantum gravity. It is important to note that these mini BHs are significantly different from classical GR BHs:
mini BHs are higher--dimensional and "stringy" at the reduced Planck scale i.e.\ more complex \cite{Dimopoulos2001}.

The lifetime of mini BHs is quite short e.g.\ a BH with mass of 3 TeV lives only $\sim 10^{-24}$ s \cite{Chamblin2004}. 
These authors also estimate that in lead--lead collisions at LHC the absorbtion rate for elementary particles is much 
smaller than the evaporation rate so that mini BHs can hardly absorb their surroundings.

The decay rate suggests that Hawking evaporation of mini BHs is rather something like an explosion. If the scenario outlined
here should be in fact verified at LHC then it is justified to entitle this as eruptive verification method of a BH.

\subsection{Obscurative methods}
%
These methods make use of the blackness of the BH. The method's name is based on the Latin word for darkness:
\textit{obscuratio}. If there is something bright in the environment the BH emerges in the foreground. In BH 
physics the luminous background source may be the accretion flow that streams like a swirl into the hole or 
it may be the cosmic microwave background (CMB).

%
The lapse function $\alpha$ has been presented in Eq. (\ref{eq:lapse_K}). This is the dominant factor producing the 
blackness. Spectral flux is reduced by higher powers of the $g$--factor which satisfies $g\propto\alpha$. The lapse
function obeys $\alpha\propto\Delta$, where $\Delta$ denotes the horizon function. $\Delta=0$ sets both horizons 
of a Kerr BH. If $\Delta\mapsto 0$, then also $g\mapsto\alpha\mapsto 0$ i.e.\ spectral flux vanishes at the horizons.

\paragraph{Black spot studies} The resulting dark feature has been called 'shadow' by \cite{Falcke2000}. We think that 
this notion is somewhat misleading because BHs do not cast a shadow geometrically as in Newtonian gravity. It is strong 
space--time curvature that warps light rays that they cannot escape from the BH. As previously suggested \cite{Mueller2004b}, 
we introduce here again the notion {\em Black Spot (BS)} for this black hole feature.

\begin{figure}[!t]\begin{center}
	\rotatebox{0}{\includegraphics[width=8.0cm]{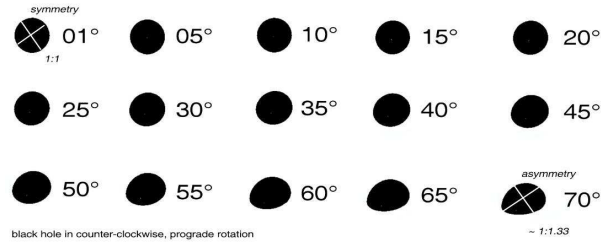}}
	\caption{A sequence of apparent shapes of the BH horizon surrounded by a flat accretion disk -- in the present 
	work called {\em Black Spots (BS)}. The study illustrates an extreme Kerr BH, $a\simeq M$, (counter--clockwise 
	rotation) as seen under different inclination angles. The shapes are numerical results from relativistic ray 
	tracing. The BS is successively deformed with increasing inclination angle. The ratio of the crossing lines can 
	be used to parametrize the deformation: At low inclination the ratio is 1:1 and the BS is more or less symmetric 
	({\em upper left}); at high inclination this ratio deviates significantly to become 1:1.3 and even more 
	({\em lower right}).} \label{fig:GBSseq}
\end{center}\end{figure}

Eq. (\ref{eq:horiz}) suggests that the intrinsic shape of the event horizon is spherically symmetric since there is 
no angle dependence. However, this radius is not an invariant. As observed from infinity the appearance of the event 
horizon depends strongly on inclination angle of the observer and on rotational state of the BH. This was first shown 
in the work by \cite{Cunningham1973} and can be confirmed by modern relativistic ray tracing simulations.
In Fig.~\ref{fig:GBSseq} a sequence is shown that proves the deformation of Kerr horizons in the middle of an accretion 
disk with variable inclination angles. These deformations can be used to determine BH properties by observations. However, 
theoretical studies show that some uncertainties are involved because the BS shapes are model--dependent e.g.\
whether or not an accretion disk is present or whether or not the accretion disk is geometrically thin 
\cite{Mueller2004b, Takahashi2004}. A detailed study of these models allows in principle for constraining inclination angle 
of the BH's rotation axis to the observer, BH spin, and BH mass if the distance of the BH candidate is known. Interferometric 
techniques especially in the radio band are supposed to allow this kind of measurements in the near future 
\cite{Falcke2000, Krichbaum2005, Krichbaum2006a}. 

%
%
%

Krichbaum et al. also suggested good candidate objects with a realistic chance to observe the BS soon 
with mm--VLBI techniques, namely the SMBH at Sgr A* and M87. The parameters for Sgr A* are 8 kpc distance and a SMBH
of mass $3\times 10^6\,{\rm M}_\odot$. M87 is with 18.7 Mpc more distant than Sgr A* but the putative SMBH is also by a factor
of thousand more massive, $3\times 10^9\,{\rm M}_\odot$, so that the BS is bigger in size, see Eq. (\ref{eq:app_bh_size}). 
   
%
\begin{figure}[!t]\begin{center}
	\rotatebox{0}{\includegraphics[width=6.0cm]{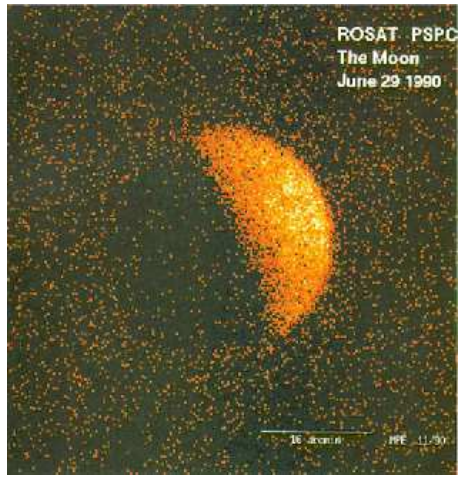}}
 	\caption{X--ray image of the sunlit half Moon taken by the German X--ray observatory ROSAT 
	in 1990 \cite{Schmitt1991}.} \label{fig:moon}
\end{center}\end{figure}
Let us now turn to the second possibility i.e.\ that the CMB acts as ambient background source \cite{Carter2006}. A 
nice analogy to this idea is shown in Fig.~\ref{fig:moon}: This is ROSAT's view of the half Moon taken in 1990 
\cite{Schmitt1991}. 
The X--ray background emerges around the black half of the Moon's disk on the 
left--hand side and outlines together with the bright right--hand side the Moon's total silhouette. 

\begin{figure}[!t]\begin{center}
	\rotatebox{0}{\includegraphics[width=6.0cm]{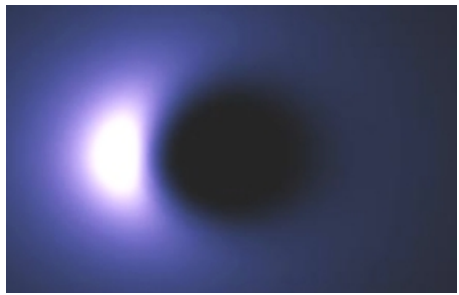}}
 	\caption{Luminous counter--clockwisely rotating accretion disk around an extremal Kerr BH. The 
	inclination angle amounts to $40^\circ$. The colour--coded intensity grows from black over blue 
	to white. The bright beaming spot is close to the BH.} \label{fig:bhdisk}
\end{center}\end{figure}

In the future, a detector of sufficient spatial resolution and sensitivity could image a comparable snapshot 
of a BH immersed in ambient radiation. This technique will deliver the first photograph of a BH and should look 
like  Fig.~\ref{fig:bhdisk}. The radiation may come from accretion flow -- or even more spectacular from the CMB. 
Using the CMB this opens the possibility to pinpoint any BH, whether or not it 
is accreting. Inactive, dormant and isolated BHs could thereby be detected! 

%

\subsection{Aberrative methods} \label{sec:aberr}
The aberrative verification technique is associated with the fact that BHs cause strong 
gravitational lensing effects. This means that the visual appearance of an object close to a BH
in generally distorted. Many examples can be found in the literature proving very strange 
simulated images e.g.\ the shape of an accretion disk around a BH. Pioneering work has been done 
by \cite{Bardeen1973, Luminet1979}. The knowledge resulting from such simulations can be used to constrain 
BH parameters. However, it must be stated that the current spatial resolution of telescopes is 
still not sufficient enough to image the warped appearance. State of the art is that astronomers use 
two alternatives to analyse images that are distorted by a BH candidate: Either they use photometric 
methods by measuring light amplification of background sources from lensing BHs in the foreground, 
\begin{figure}[!t]\begin{center}
	\rotatebox{0}{\includegraphics[width=8.0cm]{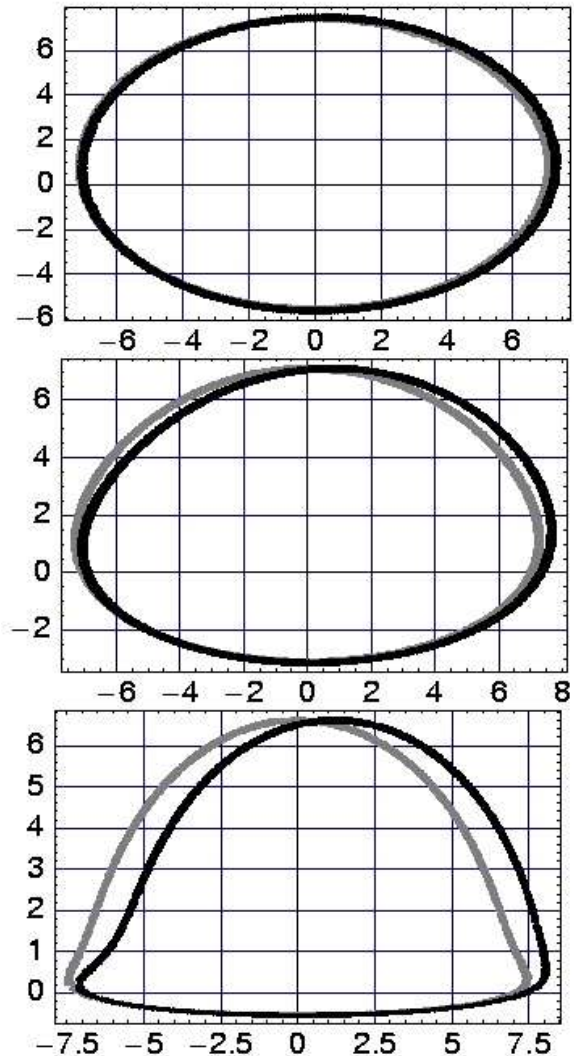}}
 	\caption{Apparent shapes of circular orbits located $r\simeq 6 \ \mathrm{r_g} $ around a 
	Schwarzschild black hole, $a=0.001$ (\textit{gray}), and a Kerr BH, $a=0.998$ 
	(\textit{black}). The apparent position of the BH is at the origin of the observers 
	screen in $(0;\;0)$. From top to bottom the inclination angles of the orbital 
	plane to the observer are $30^\circ$, $60^\circ$ and $85^\circ$. At highest inclinations a 
	clear discrepancy between a non--rotating and a rotating BH is visible.} \label{fig:distorb}
\end{center}\end{figure}
or astronomers benefit from spectroscopic methods i.e.\ the total emission over relativistically distorted 
images is measured (see Sec.~\ref{sec:spec}). 

\paragraph{Lensing of background point sources} Under special circumstances i.e.\ 
background point source, BH in the foreground, and observer are collinear characteristic emission features known as 
Einstein rings occur \cite{Einstein1936}. This phenomenon has been observed for other lens objects such as 
microlenses -- so called MACHOs i.e.\ massive compact halo objects, most probably brown dwarfs, even planets -- in 
the Galaxy \cite{Gaudi2004} and for heavy dark matter galaxy clusters that lens the background emission of distant 
galaxies \cite{Pello2004}. 
The corresponding light curve is very symmetric and can be easily selected in the zoo of observed light curves. 
However, there is to date no single observation that has proven a BH only by lensing.
But in principle, such a technique is possible. The power of that technique has been successfully tested 
in cosmology for heavy dark matter lenses, see e.g.\ the Abell catalogue.
%
%
%
%

\paragraph{Lensing of orbital shapes} Let us now turn to relativistically distorted images of tight 
orbits around BHs. We render these images by relativistic ray tracing techniques on the Kerr geometry.
The code solves the equation of motion for null geodesics in Kerr space--time i.e.\ the geodesics 
equations by using elliptical integrals \cite{Fanton1997, Mueller2004a}. All renderings presented 
here are based on the physical parameter set $\{a/M,i,r_\mathrm{in},r_\mathrm{out}\}$ where $M$ 
denotes BH mass, $a$ the Kerr parameter i.e.\ specific angular momentum of the BH, $i$ is the inclination 
angle of the orbital plane relative to the distant observer in units of degrees ($^\circ$), $r_\mathrm{in}$ 
is the inner and $r_\mathrm{out}$ the outer radius of the orbit in units of gravitational radii, 
$r_\mathrm{g}$. All results only show images of first order i.e.\ higher order images from light 
bending and returning radiation are neglected. 

Fig.~\ref{fig:distorb} shows a comparison of apparent orbit shapes around a non--rotating 
(Schwarzschild, $a=0.001\,M$, \textit{gray}) vs. a rapidly rotating BH (Kerr, $a=0.998\,M$, 
\textit{black}) as viewed by a distant observer at infinity. Each orbit is located at the ISCO, 
$r\simeq 6 \ \mathrm{r_g}$, of the Schwarzschild geometry. The variable parameter in this study is 
the inclination angle of the orbital plane that cycles over $i=30,\,60,\,85^\circ$ from top to bottom. We 
conclude that at low inclinations i.e.\ nearly face--on situations the observer cannot distinguish a 
Schwarzschild from a Kerr BH, unless he observes a \textit{stable} Keplerian orbit obeying 
$r_\mathrm{orb}\lesssim 6 \ \mathrm{r_g}$ indicating stable orbits around a Kerr black 
hole. In contrast, at high inclinations i.e.\ nearly edge--on situations the observer clearly
detects different apparent orbit shapes depending on BH spin. From the apparent orbits at the sky 
we crudely estimate the maximum relative spatial offset to be $\sim 0.08$ in the undermost example 
shown in Fig.~\ref{fig:distorb}.
%
%
%
For a 3.6--million--solar--mass BH in 7.6 kpc distance the apparent size of the orbital 
diameter, $\theta_\mathrm{orb}$, amounts to 56 $\mu$as (cf. Eq. (\ref{eq:app_bh_size}) with 
$\theta_\mathrm{orb}=3\,\theta_\mathrm{BH}$ because $r_\mathrm{orb} = 3\,R_\mathrm{S}$). Hence the 
deviation between Schwarzschild and Kerr at $i=85^\circ$ can be found to be only $\sim$4 $\mu$as. 
This is feasible by mm--VLBI techniques within the next years.
%
%
%
%
%

Let us consider also two BHXBs in contrast, Cyg X--1 satisfying $M\simeq 10\,{\rm M}_\odot$ and 
$d\simeq 2.2$ kpc as well as XTE J1118+480 with $M\simeq 7\,{\rm M}_\odot$ and 
$d\simeq 1.8$ kpc. These are the two spatially closest stellar BHs to Earth.
With Eq. (\ref{eq:app_bh_size}) it is found that the apparent sizes of these BHs
amount only to $\simeq 10^{-3}\,\mu$as in both cases (nas--scale!). This is by far out 
of current astronomical detectability.

We add here that as already shown in pioneering work strong gravitational lensing
emerges only at high inclination angles. Further, Fig.~\ref{fig:distorb} proves that 
orbit shapes become asymmetric in case of rotating BHs. This is the signature of 
frame--dragging causing matter and light to co--rotate with the space--time. 
%

Imaging of objects in the BH vicinity is characterised by relativistic blurring.
A more sophisticated example is presented in \cite{Paumard2005} resulting from 3D relativistic
ray tracing including also multiple images of higher orders (returning radiation). The 
complete orbital trajectory ressembles to a cardioid in this case where a Schwarzschild 
BH and moderate inclination of $45^\circ$ have been assumed.

Further examples with ray traced accretion flows are based on pseudo--Newtonian MHD \cite{Armitage2003} 
or GRMHD simulations \cite{Schnittman2006}. These turbulent accretion flows demonstrate that
the aberrative structure becomes more complicated.
%
%
\subsection{Temporal methods}
%
%
%
Temporal methods are based on the fact that gravity also causes time dilatation effects.
This has been verified by terrestrial experiments (Gravity Probe--A). Strong gravity from 
BHs is responsible for significant time dilatation effects as discussed at the formula for 
the lapse function on the Kerr geometry, Eq. (\ref{eq:lapse_K}). Thereby, lapse is defined 
as $\alpha=d\tau/dt$ with eigenzeit interval $d\tau$ and coordinate time interval $dt$.
Together with the gravitational redshift effect an external observer states that "an 
infalling luminous clock becomes red and faint when approaching the hole combined with a 
slow--down of the ticks". For classical GR BHs the clock stops ticking at the horizon ("freezing 
effect"). However, this is principally not observable due to the fact that it appears black for 
external observers at infinity. Additionally, tidal forces will stretch and squeeze the clock. 
The tidal damage is extraordinary strong for stellar BHs rather than for their supermassive 
counterparts (see Sec.~\ref{sec:eru}).

The idea for the temporal verification method is that a suitable 'clock', a periodical event 
in the vicinity of a BH serves as a BH indicator. Time dilation effects e.g.\ occur in light 
curves that originate from orbiters close to a BH. First calculations of the observed light 
curve of a star orbiting a Kerr BH were performed in the Seventies \cite{Cunningham1973}.

Giant and rapid X--ray variabilities have been reported for a narrow--line Seyfert--1 galaxy 
\cite{Boller1997}. These observations are consistent with strong relativistic effects that
have impact onto the X--ray light curve. 

Recently, it was shown that the modulation of a transient, redshifted iron K line in the 
Seyfert galaxy NGC~3516 can be explained by an orbiting spot on the accretion disk that is 
illuminated by a corotating flare \cite{Iwasawa2004}. Flare position and orbital timescale
have been used to estimate the SMBH mass. 

Hence in principle, such observations constrain the Kerr BH parameters. The computations by 
Cunningham \& Bardeen suggest that the the orbit radius can be determined from the light 
curve and this radius would also constrain BH mass and spin e.g.\ a Kerr BH again is signaled 
by $r_{\rm orb}< r_{\rm ISCO}(a=0)=6\,r_{\rm g}$. They were able to demonstrate that the released 
energy flux is sharply peaked in time if the orbit is close to the event horizon.
%
%
%

\subsection{Gravitational wave--induced methods}
%
%
%

As anticipated in the beginning of this Sec.~\ref{sec:detec} gravitational waves (GWs) provide the \textit{only direct} 
method for detecting BHs. Electromagnetic waves cannot serve a fiducial tracers \cite{Abramowicz2002}. It has been
demonstrated that event horizons have a characteristic imprint onto GW waveforms \cite{Ryan1995, Hughes2001}. The idea
in both proposals is that the inspiral of a compact object onto a central massive body can be used to map the
space--time. However, this appealing concept is plagued by the two main problems of GW astronomy: first, GWs have not been
detected directly so that astronomers have not these waveforms at their disposal; second, there is a source confusion problem 
in GW astronomy i.e.\ it will be hard to pinpoint the GW emitter. Another more specific problem is that e.g.\ Kerr space--times
and non--Kerr space--times could be confused \cite{Glampedakis2006}. Nevertheless, we feel that within the next years GW 
astronomy will overcome these rather technical difficulties. 

As proposed recently \cite{Berti2006}, GW waveforms (once detected) offer the opportunity to count the hairs and to rule 
out e.g.\ SMBH or boson star.
%
%
%
%
%
%

\section{Discussion}
\paragraph{What do we observe?}
Observations reconcile in many cases with classical BH solutions that are endowed with only 
two hairs: mass $M$ and spin $a$. At least in one case, alternatives to the SMBH are ruled out
i.e.\ in the Galactic Centre \cite{Genzel1997, Schoedel2002}. In other cases, there is few 
space for alternatives because rapid space--time rotation proves to be an essential ingredient 
in BHXB, GRB, and AGN physics. From this perspective, there is evidence for Kerr BHs.

However, a convincing proof consists in verifying two features of a classical BH: the event 
horizon and the intrinsic singularity. Both have not been observed by astronomical methods, yet! 
The problem is that the event horizon and its direct surroundings are too faint. The event horizon 
of classical GR BHs has \textit{per se} vanishing luminosity as discussed in Sec.~\ref{sec:black}. 
There is a controversial debate if black hole event horizons are observable \cite{Remillard2005} 
or not \cite{Abramowicz2002}. 
The ADAF argument can be exploited to distinguish neutron stars from BH candidates in X--ray binaries
\cite{Narayan1997}. However, this argument only proves that it is \textit{something different} from a
neutron star but it \textit{does not prove} event horizons. We argue that it is not possible to rule out
BH alternatives such as Gravastar or Holostar, both \textit{without} event horizons (see below) by 
this ADAF argument. 

We further stress here a temporal aspect of BHs that may be regarded as alternative interpretation of 
gravitational redshift: BH event horizons are in the future of external observers! Looking at the Penrose 
diagrams of classical BHs the intrinsic singularity has the character of a future null infinity. Therefore, 
an external observer located at asymptotical flatness does \textit{not} share the same hypersurface 
$t=const$ with the BH singularity. In simple words: BH event horizons are in our future and hence 
\textit{in principle non--observable}. As an astronomer that uses electromagnetic waves we cannot be 
assured that we observe an event horizon or the gravitational redshifted and dark sphere of a compact 
object without event horizon.

%
\paragraph{Alternatives to the classical BH} Recently, theoretical results were presented that go 
beyond Einstein. Petri found a solution of Einstein's field equation that is called holostar 
\cite{Petri2004, Petri2006}. This new compact object has an external metric that is identical to 
the Schwarzschild solution. The holostar has a thin matter shell and an internal metric that 
satisfies a total energy density distribution $\rho\propto r^{-2}$. The interior obeys an equation 
of state with anisotropic pressure. This behaviour can nicely be interpreted in the framework of 
string theories: the spherically symmetric holostar core consists of radial strings. Interestingly, 
holostars are lacking both, event horizon and curvature singularity.

Based on string theories, a BH alternative was proposed called fuzzball \cite{Mathur2004, Mathur2005, Mathur2006}.
The interior of this compact dark object consists of strings, too.

Another alternative to the Schwarzschild BH is the gravastar \cite{Mottola2002}. This solution 
exhibits an external Schwarzschild metric, too. The transition region consists of a thin matter shell 
filled with an ultrarelativistic plasma. Unlike holostars, gravastars are endowed with a dark energy 
core (de Sitter space) as internal metric. Unfortunately, the original gravastar model with isotropic 
internal pressure is unstable \cite{Vigelius2004}. But is it was shown that gravastars become stable 
and more compact with anisotropic pressure \cite{Cattoen2005}. Meanwhile, it has been demonstrated 
that the internal de Sitter metric can be replaced by other forms of dark energy 
\cite{Vigelius2004, Bilic2006a}. 

The properties and stability of these new BH--like solutions are matter of current research. If all 
these solutions turn out to be stable, astrophysicists have a serious problem: how should one single 
out one of these competing static solutions for a given Schwarzschild BH candidate? To date, the only 
possibility is that rotation of the BH candidate can be measured astronomically. Then, there is no 
alternative to the classical Kerr BH. But if researchers will find rotating generalizations of holostars,
fuzzballs, and gravastars this might lead BH astrophysics into a crisis.

\paragraph{Developments in loop quantum gravity} Besides, classical BH physics is currently confronted 
with new clues that follow from the framework of quantum geometry. Considering a simple toy model for the 
gravitational collapse of a scalar field, effects of loop quantum gravity cause a bounce in the course of 
the collapse. "Loopy effects" in granular space--time develop a negative pressure that drives strong mass
outflows. Finally, this prevents from building--up a curvature singularity. Detailed research is in progress 
but it is indicated that this effect might remove the classical BH singularity \cite{Bojowald2005, Goswami2006}.
%
%
%
%

\section{Conclusions}
Kinematical methods are very robust and deliver precise parameter measurements for both, BH mass and 
BH spin. All kinematical methods are standard in astronomy to get insights into BH properties.
We guess that reverberation mapping techniques will become more and more important especially in X--rays
because several high--precision instruments such as eROSITA, XEUS and Constellation--X are underway.

Meanwhile, spectro--relativistic verification methods such as broad iron K fluorescence lines represent 
a standard diagostic tool to study innermost accretion flow and BH. Unfortunately, prominent Fe K$\alpha$ 
lines do only occur in a fraction of all BH candidates.

Accretive methods give reliable estimates for BH mass and spin if one considers Eddington's argument.
Promising investigations involve deep surveys that give rise for average properties of
SMBHs in the field. For the future, detailed mappings of the structure around BHs with VLBI are supposed 
to deliver valuable input on the BH properties if compared to supercomputer simulations.

GRB detections as an eruptive method signal clearly the existence of a BH but only give vague
parameters for mass and spin. This is better for stellar tidal disruptions, however these events 
are too rare. The most promising upcoming eruptive method is based on decay by Hawking radiation.
But in this case the BHs have particle size. The detecability is quite uncertain because the underlying
theory offers many variants -- the serious point is that the existence of spatial extra dimensions
is questionable so far.

By means of obscurative methods the Black Spot (BS) can be measured. This becomes feasible with
mm--VLBI observations within the next few years. In principle, the BS shape allows also for
determining both, BH mass and BH angular momentum, because a higher mass results in a larger BS 
size and rotation strongly deforms the BS shape. More theoretical investigations should be done
to have a model--dependent BS gallery that can be compared to observations.

From aberrative verification methods the rotational state of
a BH can be determined from the shape of intrinsically circular orbits i.e.\ if data of a full 
revolution of a close orbiter are available. 

Temporal BH verification methods are not very common in astronomy. One possible reason for that 
might be that a suitable time sequence needs observations with longer exposure times -- this is 
expensive. Mostly, astronomers prefer other methods that rather involve spectra than light curves.

Gravitational wave astronomy is a very active research field. The existence of GWs have been documented
indirectly at the famous Hulse--Taylor pulsar. BH astrophysicists have to be patient until GW detectors
provide the first direct source detection. Then, this revolutionary new technique allows us to
probe BH space--times very precisely. As summarized here, characteristic waveforms are supposed to supply 
firsthand informations on event horizons -- the key feature of BHs. 

Whatsoever the true nature of the compact dark objects may be, mass-energy curves space--time. Null geodesics 
follow the curved space--time due to the metric interpretation of Gravity. Therefore, compact masses capture 
light. Indeed, astronomers found an impressive number of compact dark masses where BHs fit nicely in. 
These BH candidates have been detected in a broad mass range -- from the stellar up to the supermassive regime 
i.e.\ BHs with a few up to several $10^{9}\,{\rm M}_\odot$.

Very interesting ongoing research is presented in the discussions. However, strong astronomical clues that
hint for these new issues are still lacking. So far, the classical Kerr solution belonging to only a two--parameter
familiy proves very powerful to explain astronomical observations.

The next important breakthrough is expected to come from gravitational wave astronomy and we will see if the
the classical BH will succeed or fail. 

\acknowledgments{I thank the organizers of the Dubrovnik summer school for invitation and an excellent 
organisation of the whole meeting. I thank Neven Bili\' c, Rudjer Bo\v skovi\' c Institute in Zagreb (Croatia) for 
inspiring and fruitful discussions. I gratefully acknowledge the use of Fig.~\ref{fig:GCorb} by \cite{Eisenhauer2005} 
to the MPE IR group and Fig.~\ref{fig:moon} by \cite{Schmitt1991} to the MPE X--ray group.}

\begin{thebibliography}{99}
\bibitem{Abramowicz2002} 
Abramowicz, M.~A., Klu{\'z}niak, W., \& Lasota, J.-P., 2002, A\&A 396, L31
\bibitem{Adams2001} 
Adams, F.~C., Graff, D.~S., \& Richstone, D.~O., 2001, ApJ 551, L31
\bibitem{Anninos2005} 
Anninos, P., Fragile, P.~C., \& Salmonson, J.~D., 2005, ApJ 635, 723
\bibitem{Anton2006} 
Anton, L., Zanotti, O., Miralles, J.~A., et al., 2006, ApJ 637, 296
\bibitem{ArkaniHamed1998} 
Arkani-Hamed, N., Dimopoulos, S., \& Dvali, G., 1998, Phys. Lett. B 429, 263
\bibitem{Armitage2003}
Armitage, P.~J., \& Reynolds, C.~S., 2003, MNRAS 341, 1041
\bibitem{Aschenbach2004a} 
Aschenbach, B., Grosso, N., Porquet, D., \& Predehl, P., 2004, A\&A 417, 71
\bibitem{Aschenbach2004b} 
Aschenbach, B., 2004, A\&A 425, 1075
\bibitem{Balbus1991} 
Balbus, S.~A., \& Hawley, J.~F., 1991, ApJ 376, 214
\bibitem{Bardeen1973}
Bardeen, J.~M., 1973, "Rapidly rotating stars, disks, and black holes", in Black Holes, ed. C. DeWitt \& B.~S. DeWitt, Gordon and Breach, New York, 241
\bibitem{Bardeen1974} 
Bardeen, J.~M., 1974, IAU Symp.  64: Gravitational Radiation and Gravitational Collapse, 132
\bibitem{Bardeen1975} 
Bardeen, J.~M., \& Petterson, J.~A., 1975, ApJ 195, L65
\bibitem{Bender2005} 
Bender, R., Kormendy, J., Bower, G., et al., 2005, ApJ 631, 280
\bibitem{Berti2006} 
Berti, E. \& Cardoso, V., 2006 [{\tt gr-qc/0605101}]
\bibitem{Bilic2006a} 
Bili\' c, N., Tupper, G.~B., \& Viollier, R.~D., 2006, JCAP 0602, 013
\bibitem{Bilic2006b} 
Bili\' c, N., Lecture Notes to School on Particle Physics, Gravity and Cosmology, Dubrovnik, 21 Aug - 2 Sep 2006 [{\tt astro-ph/0610657}]
\bibitem{Blandford1977}
Blandford, R., \& Znajek, R., 1977, MNRAS 179, 433
\bibitem{Blandford1982}
Blandford, R.~D., \& Payne, D.~G., 1982, MNRAS 199, 883
\bibitem{Bojowald2005} 
Bojowald, M., Goswami, R., Maartens, R., \& Singh, P., 2005, Phys. Rev. Lett. 95, 091302 
\bibitem{Boller1997} 
Boller, T., Brandt, W.~N., Fabian, A.~C., \& Fink, H.~H., 1997, MNRAS 289, 393
\bibitem{Boller2006} 
Boller, T., Balestra, I., \& Kollatschny, W., 2006, accepted by A\&A [{\tt astro-ph/0612491}]
\bibitem{Bolton1972} 
Bolton, C.~T., 1972, Nature 235, 271 
\bibitem{Boyer1967}
Boyer, R.~H. \& Lindquist, R.~W., 1967, J. Math. Phys. 8, 265
\bibitem{Brusa2005} 
Brusa, M., Gilli, R., \& Comastri, A., 2005, ApJ 621, L5
\bibitem{Carr2003} 
Carr, B.~J., 2003, Lect. Notes Phy. 631, 301
\bibitem{Carter1968}
Carter, B., 1968, Phys. Rev. 174, 1559
\bibitem{Carter2006}
Carter, B., 2006, Contrib. to Encuentros Relativistas Espanoles: A Century of Relativity Theory, Oviedo, 2005 (ed. L. Mornas) [{\tt gr-qc/0604064}]
\bibitem{Cattoen2005} 
Cattoen, C., Faber, T., \& Visser, M., 2005, Class. Quantum Grav. 22, 4189 [{\tt gr-qc/0505137}]
\bibitem{Cavaglia2003} 
Cavaglia, M., 2003, IJMPA 18, 1843 [{\tt hep-ph/0210296}]
\bibitem{Chamblin2004} 
Chamblin, A., Cooper, F., \& Nayak, G.~C., 2004, Phys. Rev. D 69, 065010 [{\tt hep-ph/0301239}]
\bibitem{Colbert1999} 
Colbert, E.~J.~M., \& Mushotzky, R.~F., 1999, ApJ 519, 89
\bibitem{Cunningham1973} 
Cunningham, C.~T., \& Bardeen, J.~M., 1973, ApJ 183, 237
\bibitem{DeVillier2002} 
De~Villiers, J.-P., \& Hawley, J.~F., 2002, ApJ 577, 866
\bibitem{Dimopoulos2001} 
Dimopoulos, S., \& Landsberg, G., 2001, Phys. Rev. Lett. 87, 161602
\bibitem{Duez2005} 
Duez, M.~D., Liu, Y.~T., Shapiro, S.~L., \& Stephens, B.~C., 2005, Phys. Rev. D 72, 024028
\bibitem{Eckart1992} 
Eckart, A., Genzel, R., Krabbe, A., Hofmann, R., van der Werf, P.~P., \& Drapatz, S., 1992, Nature 355, 526
\bibitem{Eckart1996} 
Eckart, A., \& Genzel, R., 1996, Nature 383, 415
\bibitem{Eddington1924} 
Eddington, A.~S., 1924, MNRAS 84, 308
\bibitem{Einstein1936} 
Einstein, A., 1936, Science 84, 506    
\bibitem{Eisenhauer2005} 
Eisenhauer, F., Genzel, R., Alexander, T., et al., 2005, ApJ 628, 246
\bibitem{Esin1997} 
Esin, A.~A., McClintock, J.~E., \& Narayan, R., 1997, ApJ 489, 865
\bibitem{Evans1989} 
Evans, C.~R., \& Kochanek, C.~S., 1989, ApJ 346, L13
\bibitem{Fabbiano1989} 
Fabbiano, G., 1989, ARA\&A 27, 87
\bibitem{Fabian2000} 
Fabian, A.~C., Iwasawa, K., Reynolds, C.~S., \& Young, A.~J., 2000, PASP 112, 1145
\bibitem{Falcke2000} 
Falcke, H., Melia, F., \& Agol, E., 2000, ApJ 528, L13 [{\tt astro-ph/9912263}]
\bibitem{Fanton1997}
Fanton, C., Calvani, M., Felice, F. de \& Cadez, A., 1997, PASJ 49, 159
\bibitem{Ferrarese2000} 
Ferrarese, L., \& Merritt, D., 2000, ApJ 539, L9
\bibitem{Gammie2003} 
Gammie, C.~F., McKinney, J.~C., \& T$\acute{\mathrm{o}}$th, G., 2003, Astrophys. J. 589, 444
\bibitem{Gaudi2004} 
Gaudi, B.~S., \& Han, C., 2004, ApJ 611, 528
\bibitem{Gebhardt2000} 
Gebhardt, K., Bender, R., Bower, G., et al., 2000, ApJ 539, L13 
\bibitem{Genzel1997} 
Genzel, R., Eckart, A., Ott, T., Eisenhauer, F., 1997, MNRAS 291, 219
\bibitem{Genzel2003} 
Genzel, R., Sch{\"o}del, R., Ott, T., Eckart, A., Alexander, T., Lacombe, F., Rouan, D., \& Aschenbach, B., 2003, Nature 425, 934	
\bibitem{Gezari2006} 
Gezari, S., Martin, D.~C., Milliard, B., et al., 2006, Astrophys. J. 653, L25
\bibitem{Ghez1998} 
Ghez, A.~M., Klein, B.~L., Morris, M., \& Becklin, E.~E., 1998, ApJ 509, 678
\bibitem{Ghez2003} 
Ghez, A.~M., Salim, S., Hornstein, S.~D., Tanner, A., Lu, J.~R., Morris, M., Becklin, E.~E., \& Duchene, G., 2005, ApJ 620, 744
\bibitem{Giddings2002} 
Giddings, S.~B., \& Thomas, S., 2002, Phys. Rev. D 65, 056010
\bibitem{Glampedakis2006} 
Glampedakis, K., \& Babak, S., 2006, Class. Quantum Grav. 23, 4167
\bibitem{Goswami2006} 
Goswami, R., Joshi, P.~S., \& Singh, P., 2006, Phys. Rev. Lett. 96, 031302
\bibitem{Haehnelt2000} 
H{\"a}hnelt, M.~G., \& Kauffmann, G., 2000, MNRAS 318, L35
\bibitem{Halpern2004} 
Halpern, J.~P., Gezari, S., \& Komossa, S., 2004, Astrophys. J. 604, 572 [{\tt astro-ph/0402468}]
\bibitem{Hasinger2004}
Hasinger, G., 2004, Nucl. Phys. B (Proc. Suppl.) 132, 86
\bibitem{Hasinger2005} 
Hasinger, G., Miyaji, T., \& Schmidt, M., 2005, A\&A 441, 417
\bibitem{Hawking1975} 
Hawking, S.~W., 1975, Commun. Math. Phys. 43, 199
\bibitem{Henry2000} 
Henry, R.~C., 2000, ApJ 535, 350
\bibitem{Hills1975}
Hills, J.~G., 1975, Nature 254, 295
\bibitem{Hughes2001} 
Hughes, S.~A., 2001, Class. Quantum Grav. 18, 4067
\bibitem{Iwasawa2004} 
Iwasawa, K., Miniutti, G., \& Fabian, A.~C., 2004, MNRAS 355, 1073
\bibitem{Kerr1963} 
Kerr, R.~P., 1963, Phys. Rev. Lett. 11, 237
\bibitem{Koide1999} 
Koide, S., Shibata, K., \& Kudoh, T., 1999, ApJ 522, 727
\bibitem{Koide2002} 
Koide, S., Shibata, K., Kudoh, T., \& Meier, D.~L., 2002, Science 295, 1688
\bibitem{Kollatschny2003} 
Kollatschny, W., 2003, A\&A 412, L61
\bibitem{Komossa2004} 
Komossa, S., Halpern, J., Schartel, N., Hasinger, G., Santos-Lleo, M. \& Predehl, P., 2004, Astrophys. J. 603, L17 [{\tt astro-ph/0402468}]
\bibitem{Kondratko2004} 
Kondratko, P.~T., Greenhill, L.~J., \& Moran, J.~M., 2004, IAUS 2004, 325
\bibitem{Krichbaum2005} 
Krichbaum, T.~P., Zensus, J.~A., \& Witzel, A., 2005, AN 326, 548
\bibitem{Krichbaum2006a} 
Krichbaum, T.~P., Agudo, I., Bach, U., Witzel, A., \& Zensus, J.~A., 2006, to appear in the proceedings of 'The 8th European VLBI Network 
Symposium on New Developments in VLBI Science and Technology', ed. A. Marecki et al., held in Torun, Poland, on Sep 26-29, 2006 [{\tt astro-ph/0611288}]
\bibitem{Krichbaum2006b}
Krichbaum, T.~P., priv. commun.
\bibitem{Krolik2005} 
Krolik, J.~H., Hawley, J.~F., \& Hirose, S., 2005, ApJ 622, 1008
\bibitem{Li2005} 
Li, L.-X., Zimmerman, E.~R., Narayan, R., \& McClintock, J.~E., 2005, ApJS 157, 335
\bibitem{Luminet1979} 
Luminet, J.-P., 1979, A\&A 75, 228
\bibitem{LyndenBell1969} 
Lynden-Bell, D., 1969, Nature 223, 690
\bibitem{LyndenBell1971} 
Lynden-Bell, D. \& Rees, M.~J., 1971, MNRAS 152, 461
\bibitem{MacFadyen1999} 
MacFadyen, A.~I., \& Woosley, S.~E., 1999, ApJ 524, 262
\bibitem{Magorrian1998} 
Magorrian, J., Tremaine, S., Richstone, D., et al., 1998, Astrophys. J. 115, 2285
\bibitem{Magorrian1999} 
Magorrian, J., \& Tremaine, S., 1999, MNRAS 309, 447
\bibitem{Mathur2004} 
Mathur, S.~D., 2005, ~OHSTPY-HEP-T-04-001 [{\tt hep-th/0401115}]
\bibitem{Mathur2005} 
Mathur, S.~D., 2005, Fortsch. Phys. 53, 793 [{\tt hep-th/0502050}]
\bibitem{Mathur2006} 
Mathur, S.~D., 2006, Class. Quantum Grav. 23, R115 [{\tt hep-th/0510180}]
\bibitem{McHardy2006} 
McHardy, I.~M., Koerding, E., Knigge, C., Uttley, P., \& Fender, R.~P., 2006, Nature 444, 730
\bibitem{McKinney2006} 
McKinney, J.~C., 2006, MNRAS 368, 1561
\bibitem{McLure2001} 
McLure, R.~J., \& Dunlop, J.~S., 2001, MNRAS 327, 199
\bibitem{Merloni2004} 
Merloni, A., 2004, MNRAS 353, 1035
\bibitem{Meszaros1997} 
Meszaros, P., \& Rees, M.~J., 1997, ApJ 482, L29
\bibitem{Meszaros2002} 
Meszaros, P., 2002, ARA\&A 40, 137
\bibitem{Miller2006a} 
Miller, J.~M., Raymon, J., Fabian, A.~C., et al., 2006, Nature 441, 953
\bibitem{Miller2006b} 
Miller, J.~M., 2006, AN 999, No.~88, 789 [{\tt astro-ph/0609447}]
\bibitem{Mineo2000} 
Mineo, T., Fiore, F., Laor, A., et al., 2000, A\&A 359, 471
\bibitem{Misner1973} 
Misner, W., Thorne, K.~S., \& Wheeler, J.~A., 1973, Gravitation, Freeman San Francisco
\bibitem{Mottola2002} 
Mottola, E., \& Mazur, P.~O., 2002, APS Meeting Abstracts, 12011
\bibitem{Mueller2004a} 
M{\"u}ller, A., \& Camenzind, M., 2004, A\&A 413, 861
\bibitem{Mueller2004b} 
M{\"u}ller, A., 2004, Ph.D.~Thesis: Black hole astrophysics: Magnetohydrodynamics on the Kerr geometry, Landessternwarte Heidelberg, Germany
\bibitem{Mueller2006} 
M{\"u}ller, A., \& Wold, M., 2006, A\&A 457, 485
\bibitem{Narayan1992} 
Narayan, R., Paczynski, B., \& Piran, T., 1992, ApJ 395, L83
\bibitem{Narayan1994}
Narayan, R., \& Yi, L. 1994, ApJ 428, L13
\bibitem{Narayan1997} 
Narayan, R., Garcia, M.~R., \& McClintock, J.~E., 1997, ApJ 478, L79
\bibitem{Oppenheimer1939} 
Oppenheimer, J.~R. \& Snyder, H., 1939, Phys. Rev. 56, 455
\bibitem{Paumard2005} 
Paumard, T., Perrin, G., Eckart, A., et al., 2005, AN 326, 568
\bibitem{Pello2004} 
Pell{\'o}, R., Schaerer, D., Richard, J., Le Borgne, J.-F., \& Kneib, J.-P., 2004, A\&A 416, L35
\bibitem{Penrose1969} 
Penrose, R., 1969, Riv. Nuovo Cim. 1, 252
\bibitem{Petri2004} 
Petri, M., 2004 [{\tt gr-qc/0405007}]
\bibitem{Petri2006} 
Petri, M., 2006, submitted to IJMPE 
\bibitem{Pietsch2002} 
Pietsch, W., \& Read, A.~M., 2002, A\&A 384, 793
\bibitem{Porquet2004} 
Porquet, D., Reeves, J.~N., O'Brien, P., \& Brinkmann, W., 2004, A\&A 422, 85
\bibitem{Remillard2005} 
Remillard, R.~A., Lin, D., Cooper, R.~L., \& Narayan, R., 2005, AAS Meeting Abstracts 207
\bibitem{Ruffert1995} 
Ruffert, M., Janka, H.-Th., \& Schafer, G., 1995, Ap\&SS 231, 423
\bibitem{Ryan1995} 
Ryan, F.~D., 1995, Phys. Rev. D 52, 5707
\bibitem{Salpeter1964} 
Salpeter, E.~E., 1964, ApJ 140, 796
\bibitem{Schoedel2002} 
Sch{\"o}del, R., Ott, T., Genzel, R., et al., 2002, Nature 419, 694
\bibitem{Schmitt1991}
Schmitt, J.~H.~M.~M., Snowden, S.~L., Aschenbach, B., Hasinger, G., Pfeffermann, E., Predehl, P. \& Tr{\"u}mper, J., 1991, Nature 349, 583
\bibitem{Schnittman2006} 
Schnittman, J.~D., Krolik, J.~H., \& Hawley, J.~F., 2006, ApJ 651, 1031
\bibitem{Schwarzschild1916} 
Schwarzschild, K., 1916, Sitzber. Deut. Akad. Wiss. Berlin 189
\bibitem{Shapiro2005} 
Shapiro, S.~L., 2005, ApJ 620, 59
\bibitem{Shields2003} 
Shields, G.~A., Gebhardt, K., Salviander, S., et al., 2003, ApJ 583, 124
\bibitem{Shakura1973}
Shakura, N.~L., \& Sunyaev, R.~A., 1973, A\&A 24, 337
\bibitem{Streblyanska2005} 
Streblyanska, A., Hasinger, G., Finoguenov, A., Barcons, X., Mateos, S., \& Fabian, A.~C., 2005, A\&A 432, 395
\bibitem{Takahashi2004} 
Takahashi, R., 2004, ApJ 611, 996
\bibitem{Tanaka1995}
Tanaka, Y., Nandra, K., Fabian, A.C., et al., 1995, Nature 375, 659
\bibitem{Thorne1974} 
Thorne, K.~S., 1974, ApJ 191, 507
\bibitem{Thorne1986} 
Thorne, K.~S., Price, R.~H., \& MacDonald, D.~A., 1986, Black holes: The membrane paradigm, Yale University Press, New Haven and London
\bibitem{Tremaine2002} 
Tremaine, S., Gebhardt, K., Bender, R., et al., 2002, ApJ 574, 740
\bibitem{vanderKlis2000} 
van der Klis, M., 2000, ARA\&A 38, 717
\bibitem{vanParadijs2000} 
van Paradijs, J., Kouveliotou, C., \& Wijers, R.~A.~M.~J., 2000, ARA\&A 38, 379
\bibitem{Vigelius2004}
Vigelius, M., 2004, diploma~thesis: Structure and stability of Gravastars, Landessternwarte Heidelberg, Germany
\bibitem{Volonteri2003} 
Volonteri, M., Haardt, F., \& Madau, P., 2003, ApJ 582, 559
\bibitem{Volonteri2005} 
Volonteri, M., Madau, P., Quataert, E., \& Rees, M.~J., 2005, ApJ 620, 69
\bibitem{Wandel1999} 
Wandel, A., Peterson, B.~M., \& Malkan, M.~A., 1999, ApJ 526, 579
\bibitem{Wyithe2006} 
Wyithe, J.~S.~B., 2006, MNRAS 365, 1082
\bibitem{Zeldovich1964}
Zel'dovich, Y.~B. \& Novikov, I.~D., 1964, Sov. Phys. Dokl. 158, 811
\end{thebibliography}
\end{document}